\documentclass[aip,graphicx]{revtex4-1}
\usepackage{physics}
\usepackage{amsmath}
\usepackage{mathtools}
\usepackage{amssymb}
\usepackage[dvipsnames]{xcolor}
\usepackage{graphicx}
\usepackage{tabularx}
\usepackage{caption}
\usepackage{subcaption}
\usepackage{dcolumn}
\usepackage{times}
\usepackage{comment}
\usepackage{ulem}

\draft

\begin{document}

\title{Enhancing the Accuracy of Density Functional Tight Binding Models Through ChIMES Many-body Interaction Potentials}

\author{Nir Goldman}
\affiliation{Physical and Life Sciences Directorate,
             Lawrence Livermore National Laboratory, Livermore, CA 94550 USA}
\email{ngoldman@llnl.gov}
\affiliation{Department of Chemical Engineering, University of California, Davis, California 95616, United States}

\author{Laurence E. Fried}
\affiliation{Physical and Life Sciences Directorate,
             Lawrence Livermore National Laboratory, Livermore, CA 94550 USA}
\author{Rebecca K. Lindsey}
\affiliation{Department of Chemical Engineering, University of Michigan, Ann Arbor, Michigan 48109, United States}
\author{C. Huy Pham}
\affiliation{Physical and Life Sciences Directorate,
             Lawrence Livermore National Laboratory, Livermore, CA 94550 USA}
\author{R. Dettori}
\affiliation{Physical and Life Sciences Directorate,
             Lawrence Livermore National Laboratory, Livermore, CA 94550 USA}

\date{\today}

\begin{abstract}

Semi-empirical quantum models such as Density Functional Tight Binding (DFTB) are attractive methods for obtaining quantum simulation data at longer time and length scales than possible with standard approaches. However, application of these models can require lengthy effort due to the lack of a systematic approach for their development. In this work, we discuss use of the Chebyshev Interaction Model for Efficient Simulation (ChIMES) to create rapidly parameterized DFTB models which exhibit strong transferability due to the inclusion of many-body interactions that might otherwise be inaccurate. We apply our modeling approach to silicon polymorphs and review previous work on titanium hydride. We also review creation of a general purpose DFTB/ChIMES model for organic molecules and compounds that approaches hybrid functional and coupled cluster accuracy with two orders of magnitude fewer parameters than similar neural network approaches. In all cases, DFTB/ChIMES yields similar accuracy to the underlying quantum method with orders of magnitude improvement in computational cost. Our developments provide a way to create computationally efficient and highly accurate simulations over varying extreme thermodynamic conditions, where physical and chemical properties can be difficult to interrogate directly and there is historically a significant reliance on theoretical approaches for interpretation and validation of experimental results.

\end{abstract}


\maketitle

\section{Introduction}
Atomistic calculation approaches for materials modeling can be used as an independent route to aid in new materials synthesis\cite{Irle13}, characterizing mixtures for use as fuel\cite{Sharma09,MOX_fuels}, or quantifying rates for chemical decomposition of organic materials\cite{Steele_VRDAC_2020}. These types of studies generally rely on quantum mechanical approaches such as Kohn-Sham Density Functional Theory (DFT) in order to aid in experimental interpretation and/or new materials design. In particular, DFT has been shown extensively to yield accurate descriptions of condensed phase physical and chemical data, such as the material equation of state under compressive or tensile loads\cite{Schwegler08}, heats of formation/mixture of new phases\cite{Correa06,Kroonblawd_carbon_2018}, and the energetics of chemical bond breaking and forming under reactive conditions\cite{Manaa09}. However, standard DFT is also renown for its significant computational expense and poor computational scaling (generally $\mathcal{O}(N^3)$) resulting from solving for the Kohn-Sham eigenstates. As a result, DFT molecular dynamics (MD) simulations can be limited to system sizes of hundreds of atoms for timescales of tens of picoseconds or smaller for many systems\cite{Mullen_spin_lattice_2020}. In contrast, many processes of interest have properties that can span orders of magnitude larger scales, including large-scale carbon heterocycle synthesis\cite{Kroonblawd19-NPAH}, the creation of new functional materials\cite{Kulkarni_MOF_2020}, and vacancy and interstitial defect formation and mobility in solid phases\cite{Sliwa_Ta_2018}. Thus, the need for alternate simulation approaches remains a highly active research area where the goal is to develop methods that can retain much of the accuracy of quantum approaches while yielding vastly improved computational efficiency and scaling.

In this regard, machine learning approaches for the development of interatomic atomic potentials have been an effective route for modeling materials under reactive and nonreactive conditions\cite{DeepMD,Cheng_water_2020}. For example, neural networks have been used successfully to model structural properties of catalytic materials\cite{NN_structures_Kulkarni} as well as the phase stability of high-entropy ceramics\cite{aflow_curtarolo}.  \textcolor{black}{Recent efforts have used many-body kernels to achieve computationally efficient potentials with a high degree of accuracy and transferability\cite{MACE}.} Gaussian Process Regression in the form of the Gaussian Approximation Potential (GAP) has been used for a number of materials, including silicon based materials\cite{GAP_potential_Si}. Regardless, the development of these potentials
\textcolor{black}{has proven relatively data-intensive in some cases\cite{Lilienfeld_ML_charge_2017,ANI-1ccx}. This can make it difficult for these efforts} to keep up with experimental needs particularly in the area of materials synthesis, where the number of permutations of different starting materials, thermodynamic conditions, and catalysts can be combinatorially large.

Semi-empirical quantum mechanical approaches hold promise as a middle ground for accelerated simulations with a high degree of accuracy. These methods combine approximate quantum mechanics with empirical functions to yield approaches that can achieve several orders of magnitude longer time scales in quantum MD simulations.\cite{Reed12,Reed12c} In addition, semi-empirical approaches utilize significantly fewer computational resources, allowing for ensembles of statistically independent trajectories and improved statistical sampling of desired properties.\cite{Kroonblawd_polymers_2018} These methods also often show much stronger transferability to systems and conditions outside of their training set compared to interatomic potentials, in part due to the accuracy of the approximate quantum mechanics and subsequent reduced reliance on empirical functions.\cite{Tretiak_PNAS_2022}

Density Functional Tight Binding (DFTB) is one such semi-empirical quantum mechanical method\cite{DFTB+_scc,Elstner_SE_review} that has had widespread success in modeling both gas-phase molecules\cite{Elstner_DFTB_GPR018} as well as condensed matter under inert and reactive conditions\cite{Manaa02a,Goyal_DFTB3_water_2014,Irle2022}, including extreme pressures and temperatures\cite{Goldman12b,Srinivasan13}. The DFTB total energy is derived from an expansion of the Kohn-Sham energy to either second or third-order in charge fluctuations, resulting in the following expression:

\begin{equation}
E_{\rm{DFTB}} = E_{\rm{BS}} + E_{\rm{Coul}} + E_{\rm{Rep}}.
\end{equation}

\noindent Here, $E_{\rm{BS}}$ corresponds to the band structure energy, $E_{\rm{Coul}}$ is the charge fluctuation term, and $E_{\rm{Rep}}$ is the repulsive energy. $E_{\rm{BS}}$ is calculated as a sum over occupied electronic states from the DFTB Hamiltonian. The DFTB Hamiltonian matrix elements are determined from pre-tabulated Slater-Koster tables derived from reference calculations with a minimal basis set. The onsite matrix elements are the free-atom orbital energies and the off-site terms are computed with a two-center approximation where both wavefunctions and electron density are subjected to confining potentials. $E_{\rm{Rep}}$ corresponds to ion-ion repulsions, as well as Hartree and exchange-correlation double counting terms. This term can be expressed as an empirical function where parameters are fit to reproduce high-level quantum or experimental reference data. In practice, an additional dispersion correction can be included, including those in standard use for DFT calculations\cite{Tkatchenko_dispersion,DFT-D}.  \textcolor{black}{DFTB tends to exhibit $\mathcal{O}(N^3)$ scaling due to the need to solve for the band structure eigenstates, similar to DFT. However, up to three orders of magnitude increase in computational efficiency is achieved through pretabulation of the Hamiltonian and overlap matrices as well as the minimal basis set.} DFTB has been shown to exhibit transferability across element types and diverse conditions\cite{DFTB_transferability_1, DFTB_transferability_2, Kullgren_CCS_2021} and has been applied to a broad range of materials \cite{Goldman15-fm, Goldman_TiH2, DFTB_BE_1, Irle2020, Lindsey_DNTF_2020}.

However, DFTB model development can be challenging in terms of optimizing the hyperparameters needed for the approximate quantum mechanical parts of the calculations. These include the separate confining potentials for the wavefunctions and electron density (which can be different for each angular momentum channel of an element),\cite{Irle2020} choice of second-order vs. third-order charge fluctuations for the energy expression\cite{DFTB3}, and whether to use density or potential superposition when computing the Slater-Koster tables.\cite{siband-1-1,Goldman_TiH2} The DFTB Hamiltonian tends to be highly sensitive to these options\cite{Aradi17}, and in general there does not exist an intuitive prescription for creating these models from scratch. Prediction of chemical and material properties are in turn are closely coupled to $E_\mathrm{Rep}$, which itself can be determined through optimization of any number of functional forms and data sets\cite{Goldman15-fm}. The repulsive energy is usually taken to be strictly pairwise (two-center), though in many cases greater-bodied interactions can be required\cite{Goldman12b}. Novel approaches for determination of $E_\mathrm{Rep}$ include constrained spline optimization\cite{Kullgren_CCS_2021}, neural networks\cite{Tkatchenko_DFTB-NN,Niehaus_NN_Si_2022}, and Gaussian Process Regression\cite{Margraf_GPR_2020,Margraf_GPR_2021}. \textcolor{black}{Machine learning approaches for $E_\mathrm{Rep}$ can be data-intensive\cite{Tkatchenko_DFTB-NN}, non-systematic in terms of model development\cite{Elstner_DFTB_GPR018}, and potentially difficult to regularize\cite{Tretiak_PNAS_2022}, which can pose difficulties for any method that leverages these techniques}. Thus, DFTB method development would be accelerated through a more rapid approach involving rapid $E_\mathrm{Rep}$ determination with many-body effects as an option, where DFTB hyperparameters could be screened in a timely fashion and there would be a reduced reliance on time-consuming generation of quantum simulation training sets.

In this work, we discuss our recent efforts to overcome these issues through use of the Chebyshev Interaction Model for Efficient simulation (ChIMES),\cite{Lindsey17, Lindsey_AL_2020} which can be used to determine $E_\mathrm{Rep}$ for molecular and condensed phase systems relatively quickly and with comparatively lower data requirements. ChIMES is a many-body reactive force field based on linear combinations of Chebyshev polynomials. It was initially developed for pure MD simulation (i.e., where all aspects of a quantum mechanical calculation have been mapped onto the ChIMES functional form). This has included both non-reactive and reactive materials, such as water under ambient and high pressure-temperature conditions\cite{Koziol17,Lindsey_H2O_2019}, high pressure C/O systems\cite{LCO-2020, Lindsey_CO_2020}, and detonating energetic materials\cite{Pham_HN3_2021}. DFTB/ChIMES models have been created for a wide variety of materials, including actinides and their oxides\cite{Goldman_DFTB_H_Pu,Goldman_PuO2H_2022}, titanium-based systems\cite{Goldman_TiH2}, and silicon (discussed below). Additionally, ChIMES has been used to improve the accuracy of DFTB by including many-body energies and forces through $\Delta$-learning, where ChIMES augments a pre-existing DFTB parameterization for organic materials under ambient\cite{Pham_DFTB_JPCL} and reactive conditions\cite{Lindsey_DNTF_2020}. We note that similar to other machine-learning methods\cite{Tretiak_PNAS_2022}, ChIMES can be used within any semi-empirical quantum mechanical approach. However, we choose to focus on DFTB due to its close resemblance to Kohn-Sham DFT as well as its proven accuracy for a variety of materials and conditions.

We begin with a brief discussion of the ChIMES formalism, including discussion of its functional form, \textcolor{black}{computational efficiency in terms of flops required per parameter,} and methods for optimization. Next we present some recent results on a general purpose DFTB/ChIMES model for silicon polymorphs, which has remained an outstanding issue in DFTB model development. We note that all DFTB calculations discussed within this work were performed with the DFTB$+$ code\cite{DFTB+,DFTB+_current}. We then summarize previous work on a semi-automated workflow for screening DFTB hyper-parameters and $E_\mathrm{Rep}$ determination in creating a models for TiH$_2$, a candidate hydrogen storage material with several potential uses. Finally, we review our recent results in using ChIMES to create DFTB models that approach hybrid-functional and coupled cluster accuracy for organic compounds and molecular solids. In all cases, the advantages to use of DFTB/ChIMES lies in its rapid parameterization time, small data requirements relative to other machine-learned approaches, and the relative ease with which overfitting can be prevented due to regularization within linear optimization approaches as well as the orthogonal nature of the underlying basis set.

\section{Methods}
\subsection{ChIMES Formalism}
The design philosophy behind ChIMES is based on a many-body expansion of the DFT total energy. Briefly, the DFT total energy can be thought of as a sum of contributions of clusters containing different numbers of atoms:

\begin{equation}
E_{\mathrm{DFT}} =
\sum^{n_\mathrm{a}}_{i_1}                      {}^1             \!E_{i_1} +
\sum^{n_\mathrm{a}}_{i_1>i_2}                  {}^2             \!E_{i_1 i_2} +
\sum^{n_\mathrm{a}}_{i_1>i_2>i_3}              {}^3             \!E_{i_1 i_2 i_3} +
\sum^{n_\mathrm{a}}_{i_1>i_2>i_3>i_4}           {}^4             \!E_{i_1 i_2 i_3 i_4}  \\+ \dots +
\sum^{n_\mathrm{a}}_{i_1>i_2\dots\ i_{n{_\mathrm{B}-1}}>i_{n{_\mathrm{B}}}}{}^{n_\mathrm{B}}\!E_{i_1 i_2 \dots i_n{_\mathrm{B}}}.
\label{eq:genchimes}
\end{equation}

\noindent Here, the one-body energies, ${}^1\!E_{i_1}$, correspond to the atomic energy constants, the two-body energies, ${}^2\!E_{i_1 i_2}$, to all pair-wise energies with indices \{$i_1, i_2$\}, the three-body energies, ${}^3\!E_{i_1 i_2 i_3}$, to all triplet energies with indicies \{$i_1, i_2, i_3$\}, etc., all the way up to some predetermined maximum bodiedness, $n_\mathrm{B}$.  These terms are summed over all cluster combinations within the system containing $n_\mathrm{a}$ total number of atoms.

In the ChIMES formalism, we represent each of the terms in our n-body expansion as a linear combination of Chebyshev polynomials. Chebyshev polynomials of the first kind of order $m$ are defined by the expression $T_m \left(\mathrm{cos} \, \theta \right) = \mathrm{cos}\left(m \theta\right)$, more commonly written as $T_m(x)$, where $x = \mathrm{cos}\, \theta$ and thus exists over the range $\left[-1,1\right]$. Chebyshev polynomials offer a number of distinct advantages for interpolation that bear mentioning. Chebyshev polynomials of the first kind are orthogonal with respect to the weighting function $1/\sqrt{1-x^2}$. They can be computed with a recurrence relationship and define a complete basis set, allowing for arbitrary complexity in a potential energy surface. Their orthogonality allows for simple regularization where higher-order polynomial coefficients can be set to zero without necessarily adversely affecting the quality of the optimization. Polynomial expansions with Chebyshev polynomials of the first kind will have exponentially decreasing coefficients for higher-order terms due to their monic form, helping to prevent overfitting. In addition, they yield a ``nearly optimal'' error function, where the maximum error in a polynomial expansion will closely resemble a minimax polynomial.  The derivatives of Chebyshev polynomials of the first kind are related to Chebyshev polynomials of the second kind $U_m(x)$ by the expression $\mathrm{d}T_m/\mathrm{d}x = m U_{m-1}$, where $U_m\left( \mathrm{cos} \, \theta \right) = \mathrm{sin} \left[ \left(n+1\right) \theta \right]/\mathrm{sin} \, \theta$. Chebyshev polynomials of the second kind also form an orthogonal basis set (with respect to the weighting function $\sqrt{1-x^2}$) and can also be generated via a recurrence relation. This can allow for arbitrary complexity for structural optimization or molecular dynamics calculations, where atomic forces are needed.

As a result, we can now write the two-body (2B) energy term in Equation~\ref{eq:genchimes} as the following expression:

\begin{equation}
        {}^2\!E_{i_1 i_2} = f_{\mathrm{p}}\left(r_{i_1 i_2}\right) + f^{e_{i_1} e_{i_2}}_{\mathrm{c}}\left(r_{i_1 i_2}\right)
             \sum^{\mathcal{O}_{2}}_{m=1} C_m^{e_{i_1} e_{i_2}} T_m (s^{e_{i_1} e_{i_2}}_{i_1 i_2})
        \label{eqn:FF2B}
\end{equation}

\noindent In this case, $C_{m}^{e_{i_1} e_{i_2}}$ is an optimized coefficient for the interaction between atom types $e_{i_1}$ and $e_{i_2}$, taken from the set of all possible element types, $\{\boldsymbol{e}\}$. All $C_{m}^{e_{i_1} e_{i_2}}$  are permutationally invariant. $T_{m}\left(s^{e_{i_1} e_{i_2}}_{i_1 i_2}\right)$ represents a Chebyshev polynomial of order $m$, and $s^{e_{i_1} e_{i_2}}_{i_1 i_2}$ is the pair distance transformed to occur over the interval $[-1,1]$ using a Morse-like function\cite{WHBB-2009,WHBB-2011}. For that coordinate transform, $s^{e_{i_1} e_{i_2}}_{i_1 i_2} \propto \mathrm{exp}\left(-r_{i_1 i_2}/\lambda_{e_1 e_2}\right)$ and $\lambda_{e_1 e_2}$ is an element-pair distance scaling constant, frequently set to the first peak in a radial distribution function. Further details are discussed in Ref.~\citenum{Lindsey17}. The term $f^{e_{i_1} e_{i_2}}_\mathrm{c}(r_{i_1 i_2})$ is a Tersoff cutoff function\cite{Tersoff88} which smoothly varies to zero up to a predefined maximum cutoff distance. In order to prevent sampling of $r_{i_1 i_2}$ distances below those sampled in our training set, we introduce use of a smooth repulsive penalty function $f_\mathrm{p} (r_{i_1 i_2})$ that is non-zero for distances close to the inner cutoff of the Chebyshev polynomials.

Many-body (e.g., greater than two-body) orthogonal polynomials can be created by defining a cluster of size $n$ and taking the product of the Chebyshev polynomials derived from the constituent ${n \choose 2}$ unique pairs. For example, the three-body polynomials will be products of ${3 \choose 2} = 3$ two-body polynomials. We thus write the ChIMES three-body (3B) energy as the following:

\begin{equation} 
{}^3\!E_{i_1 i_2 i_3} =       f^{e_{i_1} e_{i_2}}_{\mathrm{c}}\left(r_{i_1 i_2}\right)
                                f^{e_{i_1} e_{i_3}}_{\mathrm{c}}\left(r_{i_1 i_3}\right)
                                f^{e_{i_2} e_{i_3}}_{\mathrm{c}}\left(r_{i_2 i_3}\right)
                \sum^{\mathcal{O}_{3}}_{m=0}
                \sum^{\mathcal{O}_{3}}_{p=0}
                {\sum^{\mathcal{O}_{3}}_{q=0}}^\prime
                C^{e_{i_1} e_{i_2} e_{i_3}}_{mpq}
                                T_{m}\left(s^{e_{i_1} e_{i_2}}_{i_1 i_2}\right)
                                T_{p}\left(s^{e_{i_1} e_{i_3}}_{i_1 i_3}\right)
                                T_{q}\left(s^{e_{i_2} e_{i_3}}_{i_2 i_3}\right).
        \label{eqn:FF3B}
\end{equation}

\noindent We thus compute a triple sum for the product of the $i_1 i_2$, $i_1 i_3$, and $i_2 i_3$ pair-wise polynomials. These are computed up to a predefined order ($\mathcal{O}_{3}$) for each three-body polynomial and then multiplied by a single coefficient, $C^{e_{i_1} e_{i_2} e_{i_3}}_{mpq}$, that is permutationally invariant for each set of polynomial orders and atom types,. The primed sum in Equation~\ref{eqn:FF3B} indicates that only terms for which two or more of the $m,p,q$ polynomial powers are greater than zero are included in order to guarantee that all atoms in the cluster interact through Chebyshev polynomials.  
The expression for ${}^3\!E_{i_1 i_2 i_3}$ also contains the $f_{\mathrm{c}}$ smoothly varying cutoff functions for each constituent pair distance. The penalty function $f_p$ is not included for many-body interactions.

Higher bodied terms are included in ChIMES in a similar fashion. For example, four-body (4B) terms are often included in ChIMES optimizations\cite{Pham_HN3_2021}, where ${}^4\!E_{i_1 i_2 i_3 i_4}$ is now determined from the sum over the product of the ${4 \choose 2} = 6$ constituent pair-wise polynomials multiplied by a single permutationally invariant coefficient. In the determination of permutational invariance for an arbitrary number of bodies, it is important to realize that the atom indices and atom types are permuted, which then implies a corresponding permutation of the bond distance indices in $C$.   In practice, even higher bodied terms could be included in ChIMES, though this can lead to a combinatorially large polynomial space and hence parameter explosion that can lead to overfitting and excessive computational expense. Hence, the norm with ChIMES optimization is generally to include up to four-body terms, though DFTB/ChIMES models are often converged with up to three-body terms, only.\cite{Goldman_DFTB_H_Pu,Lindsey_DNTF_2020,Goldman_TiH2,Pham_DFTB_JPCL,Goldman_PuO2H_2022}

ChIMES bears some resemblance to the Atomic Cluster Expansion approach (ACE)\cite{Drautz_ACE1,Csanyi_ACE_organic}, where many-body interactions are represented by a product of Chebyshev polynomials and real spherical harmonics. These models also differ from ChIMES in that the underlying polynomial basis set is atom-centered (similar in spirit to an embedded atom model\cite{Fichthorn09a}) rather than using a cluster approach as we adopt here. Similarly, the spectral neighbor analysis potential (SNAP) uses bispectrum components to compute the total energy of a system as a sum over atom energies, which are expressed as a weighted sum over bispectrum components\cite{Thompson_SNAP}.

Similar to other machine learning atomic interaction potentials, ChIMES uses the method of force matching\cite{Ercolessi94} to determine the interaction parameters.
In force matching a training set of quantum simulations is generated with differing configurations.  For each configuration in the training set, the quantum mechanical
energy, atomic force, and stress (for condensed phase configurations) are calculated.  The ChIMES parameters are varied so as to minimize the error in these quantities. The determination of an appropriate training set is perhaps the most difficult part in generating an interaction potential.  The configurations in the training set
need to sample the relevant configurations of each cluster in the ChIMES model, yet should avoid highly unfavorable configurations that could be difficult
to converge with quantum theory and could require very high polynomial orders to represent with ChIMES.   A generally successful approach has been to generate configurations by sampling molecular dynamics simulations with ab initio forces.  Trajectories with multiple densities, temperature, and elemental compositions are calculated with the quantum method.  The range of densities, temperatures, and elemental compositions should reflect the intended use of the model.  
For non-DFTB ChIMES force field models, it has also proven useful to decompose condensed phase configurations into bonded clusters, so that the energy of each cluster may be determined through quantum mechanics. \textcolor{black}{This has allowed for more accurate descriptions of condensed phase chemistry, including decomposition under extreme conditions.\cite{Lindsey_DNTF_2020,Pham_HN3_2021}}. Active learning methods and cluster resampling have been essential to effectively sample unstable cluster configurations, such as those occurring at chemical transition states\cite{Lindsey_AL_2020}.  This has not been necessary for DFTB/ChIMES models, where the ChIMES force field plays a more minor role in determining the overall system energetics.

In addition, weights are required when matching to forces, energies, and stresses, due to the differing physical units and number of parameters per configuration.  Since there are 3N forces for an N atom system, but only 1 energy and 6 unique stress tensor elements, non-trivial weights for energy and stress are usually required to achieve desired levels of accuracy in these quantities.  Once weights are determined, an objective function for optimization may be defined as follows:

\begin{footnotesize}
\begin{equation}
        F_{\mathrm{obj}}  = 
        \frac{1}{N_d}
\sum_{\tau=1}^M 
\left(
\sum_{i=1}^{N_\tau}
\sum_{\alpha=1}^{3} \left(w_{\mathrm{F}}\Delta \mathrm{F}_{\tau_{\alpha_i}} \right)^{2}
+
\sum_{\alpha=1}^3 \sum_{\beta \le \alpha}\left(w_{\sigma}\Delta\sigma_{\tau_{\alpha\beta}}\right)^{2}
+
\left(w_{\mathrm{E}} \Delta E_{\tau}\right)^{2}
\right)
.
\label{eqn:rmse}
\end{equation}
\end{footnotesize}

\noindent Here, $\tau$ corresponds to a specific training set configuration, $i$ is the atomic index, and $\alpha$ and $\beta$ are the cartesian directions. 
$M$ is the total number of configurations in the training set and $N_d = 3\sum_\tau N_\tau + 7 M$ is the total number of data entries (6 stress tensor components and one energy value per configuration).  In addition, $\Delta F_{\tau_{\alpha_i}} = F^{\mathrm{ChIMES}}_{\tau_{\alpha_i}} - F^{\mathrm{DFT}}_{\tau_{\alpha_i}}$, $\Delta\sigma_{\tau_{\alpha\beta}} = \sigma^{\mathrm{ChIMES}}_{\tau_{\alpha\beta}} - \sigma^{\mathrm{DFT}}_{\tau_{\alpha\beta}}$, 
and $\Delta E_{\tau} = E^{\mathrm{ChIMES}}_{\tau} - E^{\mathrm{DFT}}_{\tau}$.  The value $w_{\mathrm{F}}$ is the weight for forces, $w_{\sigma}$ for stresses, and $w_{\mathrm{E}}$ for energies.

Optimal ChIMES parameters are determined by solving 
\begin{equation}
\frac{\partial F_{\mathrm{obj}}}{\partial C_I} = 0,
\label{eqn:optF}
\end{equation}
for all $I$, where $I$ is a combined index of the permutationally unique coefficient.  For example, for a two body interaction, $I = \{e_1, e_2, m\}$.
Optimizing $F$ is equivalent to solving the overdetermined matrix equation 
\begin{equation}
\boldsymbol{w} \boldsymbol{A} \boldsymbol{C} = \boldsymbol{w} \boldsymbol{B}.
\label{eqn:linear}
\end{equation}
The matrix $\boldsymbol{A}$ corresponds to the derivatives of the ChIMES energy, stress, or force expression with respect to the fitting coefficients.  The column vectors $\boldsymbol{C}$ and $\boldsymbol{B}$ correspond to the ChIMES coefficients to be optimized and the numerical values for the training data, respectively. The diagonal matrix $\boldsymbol{w}$ is comprised of the weights to be applied to the
elements of $\boldsymbol{B}$ and rows of $\boldsymbol{A}$. Solution to this linear least-squares problem can performed using a number of different optimization algorithms, which we discuss in more detail below.

\subsection{ChIMES optimization for $E_\mathrm{Rep}$ or $\Delta$-learning}

The ChIMES/DFTB parameter determination of $E_\mathrm{Rep}$ or $\Delta$-learning proceed in a similar fashion to that described above for a ChIMES force field. $E_\mathrm{Rep}$ training is computed by calculating DFTB forces $(F)$, stress tensor components $(\sigma)$, and possibly system energies $E_\mathrm{tot}$ for each configuration in the training set with the chosen set of Hamiltonian parameters (i.e., the set of wavefunction confining potentials $\{R_{\psi}\}$, density confining potentials $\{R_{n}\}$, use of density or potential superposition, second or third-order DFTB) with zero values for those components from $E_\mathrm{Rep}$. These ``repulsive energy free'' results are then subtracted from their corresponding DFT values, i.e.,
\begin{eqnarray}
E_{\tau}^\mathrm{Rep'} &=& E_{\tau}^{\mathrm{DFT}} - E_{\tau}^{\mathrm{QM,DFTB}} \nonumber \\
F^{\mathrm{Rep'}}_{\tau_{\alpha_i}} &=& F^{\mathrm{DFT}}_{\tau_{\alpha_i}} - F^{\mathrm{QM,DFTB}}_{\tau_{\alpha_i}} \nonumber \\
\sigma^{\mathrm{Rep'}}_{\tau_{\alpha\beta}} &=& \sigma^{\mathrm{DFT}}_{\tau_{\alpha\beta}} - \sigma^{\mathrm{QM,DFTB}}_{\tau_{\alpha\beta}}
\label{eqn:erep_training}
\end{eqnarray}
and the objective function is modified as follows:

\begin{footnotesize}
\begin{equation}
        F_{\mathrm{obj}}  = 
        \frac{1}{N_d}
\sum_{\tau=1}^M 
\left(
\sum_{i=1}^{N_\tau}
\sum_{\alpha=1}^{3} \left(w_{\mathrm{F}}\Delta \mathrm{F}^{\mathrm{Rep}}_{\tau_{\alpha_i}} \right)^{2}
+
\sum_{\alpha=1}^3 \sum_{\beta \le \alpha}\left(w_{\sigma}\Delta\sigma^{\mathrm{Rep}}_{\tau_{\alpha\beta}}\right)^{2}
+
\left(w_{\mathrm{E}} \Delta E^{\mathrm{Rep}}_{\tau}\right)^{2}
\right)
,
\label{eqn:rmserep}
\end{equation}
\end{footnotesize}
where $\Delta F^{\mathrm{Rep}}_{\tau_{\alpha_i}} = F^{\mathrm{ChIMES}}_{\tau_{\alpha_i}} - F^{\mathrm{Rep'}}_{\tau_{\alpha_i}}$, 
$\Delta\sigma^{\mathrm{Rep}}_{\tau_{\alpha\beta}} = \sigma^{\mathrm{ChIMES}}_{\tau_{\alpha\beta}} - \sigma^{\mathrm{Rep'}}_{\tau_{\alpha\beta}}$, 
and $\Delta E^{\mathrm{Rep}}_{\tau} = E^{\mathrm{ChIMES}}_{\tau} - E^{\mathrm{Rep'}}_{\tau}$.

In practice, we have used the diagonal components of the stress tensor, only (i.e., $\alpha = \beta$ in Equation~\ref{eqn:rmserep}). `QM,DFTB' in Equation~\ref{eqn:erep_training} refers to the quantum components of the DFTB calculation, i.e., only forces and stresses from $E_\mathrm{BS}$ and $E_\mathrm{Coul}$. Calculation of a $\Delta$-learning training set is identical with the exception that the quantities in Equation~\ref{eqn:rmserep} are no longer repulsive energy free but instead contain terms from the DFTB repulsive energy model of choice. 

\subsection{Linear least-squares approaches for ChIMES optimization}

The ChIMES potential is linear with respect to the fitting coefficients, which allows for use of powerful global optimization tools that are unavailable to non-linear machine-learned models.  Even though the ChIMES parameters are formally overdetermined in the optimization of $F_{\mathrm{obj}}$, in practice there are usually parameters or linear combinations of parameters that are ill-determined.  One issue is that configurations sampled from molecular dynamics simulations are often highly correlated with one another, so that there are strong correlations between the properties in differing configurations of the training set.  If certain clusters are not sampled in the training set, due to unfavorable energetics or choices made by the model developer, then parameters describing that cluster will not be determined.  For example, in developing a ChIMES potential for $\mathrm{H}_2\mathrm{O}$, the short-range interaction of three O atoms would likely not be sampled unless systems other than $\mathrm{H}_2\mathrm{O}$ (such as pure $\mathrm{O}_2$) were used in the training set.

In order to overcome this issue, some form of regularization is required to avoid unphysically large parameter magnitudes that can occur when the matrix 
$\boldsymbol{A}$ is ill-conditioned.  In our efforts, we have focused on the Singular Value Decomposition (SVD) and  Least Absolute Selection and Shrinkage Operator (LASSO) regularization methods, which we discuss briefly. Principal component analysis using SVD\cite{num_recipes} solves Equation~\ref{eqn:linear} for optimal fitting coefficients directly by computing the pseudoinverse of the generally rectangular $\boldsymbol{A}$ matrix from its eigendecomposition. This yields singular values which are the eigenvalues of the generated square matrix. The optimization process can be regularized by setting singular values with an absolute value below a given threshold to zero. In our work, we take this parameter to be $D_{max} \epsilon$, where $D_{max}$ is the maximum singular value of $\boldsymbol{A}$ and $\epsilon$ is a factor below a value of one.

LASSO\cite{lasso} is an $L^1$-norm regularization method whereby regularization is based on the sum of the absolute values of the fitting coefficients, which has the effect of shrinking a subset of parameters to zero. In this case, the objective function $F_\mathrm{obj}$ (Equation~\ref{eqn:rmse}) is minimized with the following additional penalty on parameter absolute values:
\begin{equation}
F^{\mathrm{LASSO}}_\mathrm{obj} = N_d F_{\mathrm{obj}} + 2\alpha \sum_{i=1}^{N_p} \left|C_i\right|.
\end{equation}
\noindent Here, $N_p$ is the total number of unique fitting parameters, $C_i$. The parameter $\alpha$ regularizes the magnitude of the fitting coefficients, which reduces possible overfitting.  

The LASSO method is highly studied in statistics, due to its ability to optimally select a subset of parameters that best describe the data, although we are not aware of its use in potential energy model development.  We chose LASSO for ChIMES molecular dynamics simulations, in which setting a large fraction of the parameters to zero is desirable for numerical efficiency.  For DFTB/ChIMES simulations, the ChIMES calculation is much faster than the quantum calculation, so the numerical efficiency of ChIMES is less of a concern.  Parameter selection, however, is still desirable when particular cluster configurations are poorly sampled.  LASSO will automatically set parameters corresponding to an un-sampled or poorly sampled cluster configuration to zero.  

There are a number of algorithms for the solution of the LASSO equations.  The most commonly used is coordinate descent\cite{lars2}, in which parameters are set one at a time using a computationally inexpensive updating formula.  Unfortunately, coordinate descent has poor convergence properties when the $\boldsymbol{A}$ matrix is ill conditioned.  Alternatively, the LASSO method can be implemented as a variant of Least-Angle Regression (LARS)\cite{lars1}.  We find that this
method works well for poorly conditioned $\boldsymbol{A}$.  
In the LASSO variant of LARS, all model coefficients are initialized to zero and we determine the covariate (i.e., ChIMES coefficient) most correlated to the error residual.   The fitting coefficient is then increased to minimize the error residual until a second coefficient is equally correlated, upon which it is included in the fitting (active) parameter set and both coefficients are modified simultaneously, and so on. For the LASSO variant of LARS parameters may be removed from the non-zero set under certain conditions. The process can be continued until all coefficients are included in the solution, at which point a result equivalent to ordinary least squares fitting is obtained.  
Each step of the method is a unique solution of the LASSO equations for a value of $\alpha$ that is larger than the target value.  This allows for  
analysis of the solution accuracy vs. parameter magnitude for an entire family of solutions, which we have not fully utilized in our studies to date. The numerical complexity of SVD and LARS/LASSO are similar, each requiring $\order{N_p^2 N_d}$ floating point operations\cite{num_recipes,lars1}.

The number of parameters in a ChIMES model depends sensitively on the number of elements, the polynomial order, and the bodiedness of the calculation.
We give some examples in Table~\ref{tab:params}.  \textcolor{black}{The number of ChIMES interactions to consider scales combinatorially with the bodiedness of the model and the number of elements, i.e., $\left(n \choose r\right)$  = $\left(n+r-1\right) \choose r$, where $n$ is the number of elements to be considered and $r$ is the bodiedness of the interaction (e.g., two, three, or four-body). For example, a three-element system will require evaluating up to $\left(3 \choose 2\right) = 6$ different two-body interactions, $\left(3 \choose 3\right) = 10$  three-body interactions, and $\left(3 \choose 4\right) = 15$ four-body interactions. ChIMES potentials are computed over the polynomial hypercube. This nominally implies $8^6$ coefficients for each 8th order four-body interaction, though the actual number of unique coefficients is significantly smaller due to permutational invariance.} As a rule of thumb, the SVD or LASSO/LARS algorithms can be conveniently solved for a problem with 5,000 or fewer parameters in less than on hour on a single Intel computer node using Python libraries such as Scikit-learn\cite{scikitlearn}.  

Parameter counts above 5,000 often
occur when 4B interactions are used, which are required for ChIMES force field models of complex organic material chemistry.  Typically, DFTB/ChIMES models do
not require 4B interactions, as discussed in the Results section.  For problems with greater than 5,000 parameters, we have developed a parallel code called DLARS that implements the LASSO/LARS algorithm with distributed memory and parallel rank-1 Cholesky decomposition updates used in solving for parameter values.  LASSO/LARS could be particularly advantageous for problems with a very large number of parameters, but a significantly smaller number of non-zero parameters.  For example, LASSO/LARS could automatically determine meta-parameters such as the Chebyshev orders given a desired accuracy level.  The distributed memory feature of DLARS 
supports very large training sets, in excess of 1 TB. As we discuss below, DFTB/ChIMES typically has more modest training set requirements. 

\textcolor{black}{The number of flops required to compute the ChIMES total energy can be estimated theoretically. First, each new $T_{n+1}$ polynomial value requires three flops to be computed from the $T_n$ and $T_{n-1}$ values, according to the Chebyshev recursion relation. The number of flops required to compute a single term in the energy sum can be expressed as $3 N_\mathrm{pair} + \left(N_\mathrm{pair} - 1\right) +1 = 4 N_\mathrm{pair}$. Here, $N_\mathrm{pair}$ is the number of unique distance pairs in the cluster. The factor of $\left(N_\mathrm{pair} - 1\right)$ is due the cost of multiplying the individual polynomial values together to create a many-body polynomial, and the extra factor of `$+1$' at the end of the left side of the equation corresponds to multiplying by the ChIMES coefficient. }

\textcolor{black}{If we then express the number of non-unique parameters as $N_\mathrm{cp}$ and include the cost of the coordinate transform ($F_\mathrm{Tr}$) and cutoff function ($F_\mathrm{C}$), the total computational cost to compute the energy of the entire cluster, $F_\mathrm{total}$, can be expressed as:}

 \begin{eqnarray}
 \color{black} F_\mathrm{total} & \color{black} = & \color{black} N_\mathrm{pair} \left(F_\mathrm{C}+1\right) + N_\mathrm{pair} \left(F_\mathrm{Tr}\right) + 4 N_\mathrm{cp} N_\mathrm{pair} + \left(N_\mathrm{cp}  - 1\right)  \nonumber \\ 
& \color{black} = & \color{black} N_\mathrm{pair} (F_\mathrm{C} + F_\mathrm{Tr} + 1) + N_\mathrm{cp} \left(4 N_\mathrm{pair} + 1\right) - 1
 \label{eqn:flops}
\end{eqnarray}

\noindent \textcolor{black}{Similarly, the factor of `$+1$' for the $F_\mathrm{C}$ terms on the left side of the top equation comes from multiplication of the value of cutoff function with the value of the sum of the polynomial terms times their coefficients. The factor of $4 N_\mathrm{cp} N_\mathrm{pair}$ is the cost of calculating all the terms in the polynomial sum individually, while $N_\mathrm{cp} - 1$ is the cost of adding these terms together.} 

\textcolor{black}{In summary, the computational cost of a ChIMES energy calculation can be estimated to be relatively low, with a theoretical value that varies as four flops per polynomial coefficient. In practice, the atom-pair cutoff and coordinate transform values could be precomputed for all clusters and stored, reducing computational cost even further. We note that ChIMES forces and stress tensor components are more expensive to calculate, and our code implementation may not have reached full theoretical efficiency.}

\begin{table}[!htp]
\caption{Number of parameters ($N_p$) for varying number of elements ($N_e$) and
Chebyshev polynomial order for 2, 3, and 4 bodied interactions ($\mathcal{O}_{\{2,3,4\}}$).}
\label{tab:params}
\begin{tabular}{ccccr}
$N_e$ & $\mathcal{O}_2$ & $\mathcal{O}_3$ & $\mathcal{O}_4$ & $ N_p$ \\
\hline
2 & 8 & 0 & 0 & 24 \\
2 & 8 & 8 & 0 & 794 \\
2 & 8 & 8 & 4 & 3,966 \\
2 & 8 & 8 & 8 & 184,306 \\
3 & 8 & 0 & 0 & 48 \\
3 & 8 & 8 & 0 & 2,512 \\
3 & 8 & 8 & 4 & 17,389 \\
3 & 8 & 8 & 8 & 912,085
\end{tabular}
\end{table}

Software for the development of ChIMES models, including DLARS, is publicly available at \url{https://github.com/rk-lindsey/chimes_lsq}.
Software for the evaluation of forces, stresses, and energies for ChIMES force field models is available at \url{https://github.com/rk-lindsey/chimes_calculator}.
ChIMES has been integrated into the DFTB+ program, which is available at \url{https://dftbplus.org}.

\section{Results}
\subsection{DFTB/ChIMES Models for Silicon Polymorphs}

Silicon has proven to be a significant challenge for DFTB model parameterization likely due to the fact that its different polymorphs can have different coordination numbers and nearest neighbor distances. This yields a variety of bond lengths and energies that need to be accounted for in order to obtain a single, transferable DFTB model that does not have to be specific for a given solid phase. Previous work has shown that standard two-body repulsive energies do not exhibit sufficient complexity to accurately account for several Si phases with different bonding environments,\cite{Kullgren_CCS_2021} in contrast to carbon, where multiple phases can be represented by a single two-body polynomial expansion\cite{Goldman12}. Neural network (NN) approaches have been used for the repulsive energy in order to account for many-body interactions in $E_\mathrm{Rep}$,\cite{Niehaus_NN_Si_2022} and the results are promising. NN approaches though generally require large amounts of data and can frequently optimize to local minima, potentially complicating their use. Here, we attempt to overcome this issue by creating a many-body ChIMES $E_\mathrm{Rep}$ for silicon that is transferable to a number of different Si polymorphs as well as prediction of vibrational spectra and calculation of defect formation energies.

In our work, we target two previous Si DFTB parameterizations, pbc-0-3\cite{pbc-0-3} and siband-1-1,\cite{siband-1-1} which have different strengths and weaknesses. The pbc-0-3 parameter was creating using density superposition (i.e., the quantum mechanical potential $V_\mathrm{QM}\left(\rho\right)$ was expressed as $V \left(\rho_A + \rho_B\right)$ for atoms $A$ and $B$) , which tends to be preferred due to its improved representation of chemical bonding and vibrations\cite{Goldman_TiH2}. However, d-orbital interactions were not tabulated aside from the d-orbital onsite energy, which could have ramifications for some material properties. In contrast, the siband-1-1 parameter set was specifically created with d-orbital interactions but with potential superposition (i.e., $V_\mathrm{QM}\left(\rho\right) = V \left(\rho_A\right) + V \left(\rho_B\right)$) in order to yield accurate prediction of electronic properties, including the electronic band structure of Si-containing solids. In addition, the siband-1-1 parameter set does not contain a repulsive energy of any sort, precluding its use in structural relaxation or MD simulation which severely limits its usefulness overall. 

Our goal is to thus to create new ChIMES $E_\mathrm{Rep}$ potentials for each set of Slater-Koster interaction parameters using identical DFT training data and ChIMES hyperparameters in order to compare and contrast the effectiveness of each as a possible one-fits-all  model. Calculations for our silicon DFT dataset were performed using the Vienna ab initio Simulation Package (VASP)\cite{vasp,vasp2,vasp3}, with projector-augmented wave function (PAW) pseudopotentials\cite{Bloechl94,Kresse99} and the Perdew-Burke-Ernzerhof exchange-correlation functional (PBE)\cite{Perdew:1996}. We found our results to be converged with a planewave cutoff of 500~eV, which was used in all of the calculations discussed here. We have used an electron density convergence criteria of $10^{-6}$~eV, with a force convergence of $10^{-2}$~eV/\AA\ for all geometry/cell optimizations. The Mermin functional\cite{Mermin65} smearing was set to 0.03~eV for all calculations performed in this work. The system energy and pressure was found to be converged with sampling of the Brillouin Zone with a $2 \times 2 \times 2$ Monkhorst-Pack mesh\cite{Monkhorst76} for all supercells. We then generated cold curves for each phase by isotropically expanding and contracting the simulation cell lattice. Here, we used a diamond structure supercell of 64 atoms, a bcc structure of 54 atoms, a simple cubic structure of 64 atoms, and a graphene sheet of 32 atoms. This yielded an initial set of 463 configurations for our ChIMES $E_\mathrm{Rep}$ optimization. \textcolor{black}{All of these configurations were given a weight of one, excluding the graphene cold curve data, where the energy was given a weight of two to account for the smaller number of atoms relative to the other phases, and the force components within the basal plane were given a weight of three in order to offset the negligible forces normal to the plane.}

In order to sample forces from a variety different configurations, we have also included MD data for the diamond and graphene phases, using the same number of atoms in each supercell as before. These supercells were isotropically expanded and contracted between 90\% to 110\% of the ground-state density. Each MD simulation was run for $\sim$5~picoseconds at 600~K, from which we took snapshots at fixed intervals of $\sim$200~femtoseconds for our training set. This yielded an additional 405 configurations for our ChIMES $E_\mathrm{Rep}$ determination. \textcolor{black}{All of the MD configuration data (e.g., forces, stress tensor diagonals, and energies) were given a weight of ten. This weighting had the effect of reducing some of the oscillations found in the cold curve data for expanded states (discussed below).} In all, our final training set contained a total of 838 configurations of different silicon phases. ChIMES $E_\mathrm{Rep}$ optimization was then performed using values of $r_\mathrm{min} = 2.0$ \AA\ and $r_\mathrm{max} = 4.0$ \AA. The value of $r_\mathrm{max}$ was informed in part from previous development of a neural network repulsive energy,\cite{Kullgren_CCS_2021} which resulted in a minimization of the root mean square (RMS) error in our fit. In addition, we found that a value of $r_\mathrm{max} = 4.0$ \AA\ yielded an improved description of the expanded states in our training set, where the bonded interactions between Si atoms is longer than the ground-state.

We now refer to our ChIMES model based on pbc-0-3 as pbc/ChIMES and our model based on siband-1-1 as siband/ChIMES. Both pbc/ChIMES and siband/ChIMES were created with a 2B order of 12, 3B order of 8, and a LASSO regularization parameter ($\alpha$) value of $10^{-3}$, similar to previous efforts\cite{Goldman_TiH2}. We have used the Morse coordinate transform with a value of $\lambda = 2.4$ \AA, which corresponds to the first peak in the diamond phase radial distribution function. For pbc/ChIMES, this yielded an overall RMS error of 1.44 eV/\AA\ in the forces, 0.43 GPA in the pressure, and 0.038 eV/atom in energy. The RMS errors for siband/ChIMES were slightly higher, with values of 2.22 eV/\AA\ for the forces, 0.55~GPa for the pressure, and 0.16~eV/atom for the energy. Use of a Chebyshev basis set 2B order of 16, 3B order of 12, and 4B order of 4 yielded reduction in the RMS errors of $<1$\% with similarly marginal improvement in validation quantities such as the computed defect energies. Use of a value of $\lambda = 3.0$ \AA\ also had only a small effect on the resulting model. All ChIMES/DFTB calculations were performed with self-consistent charges using similar parameters to our DFT calculations. This included charge convergence criteria of  $2.72 \times 10^{-5}$ eV ($10^{-6}$ au), a force convergence of $10^{-2}$ eV/\AA\ for all geometry optimizations, and $2 \times 2 \times 2$ k-point mesh for all calculations.

\begin{table}[!htp]
  \footnotesize
  \caption{Ground state energies relative to diamond ($\Delta E_\mathrm{diam}$) in eV/atom and nearest neighbor distances (NN) in \AA\ for the Si polymorphs considered in this work.}
  \label{tab:si_polymorphs}
\begin{tabular}{c cc| cc| cc| cc| cc| cc}
 & \multicolumn{2}{c}{diamond} & \multicolumn{2}{c}{bcc} &  \multicolumn{2}{c}{simple cubic} &  \multicolumn{2}{c}{graphene} &  \multicolumn{2}{c}{bc8} & \multicolumn{2}{c}{\textcolor{black}{$\beta$-Sn}}\\
\hline
\hline
& NN & $\Delta E_\mathrm{diam}$ & NN & $\Delta E_\mathrm{diam}$ & NN & $\Delta E_\mathrm{diam}$ & NN  & $\Delta E_\mathrm{diam}$ & NN & $\Delta E_\mathrm{diam}$ & \textcolor{black}{NN} & \textcolor{black}{$\Delta E_\mathrm{diam}$} \\
pbc/ChIMES 	 & 2.37 & 0.00 & 2.67 & 0.55 & 2.53 & 0.30 & 2.23 & 0.70 & 2.37 & 0.14 & \textcolor{black}{2.63} & \textcolor{black}{0.24} \\
siband/ChIMES & 2.36 & 0.00 & 2.65 & 0.53 & 2.54 & 0.31 & 2.26 & 0.59 & 2.39 & 0.15 & \textcolor{black}{2.47} & \textcolor{black}{0.25} \\
DFT 			 & 2.37 & 0.00 & 2.68 & 0.54 & 2.53 & 0.30 & 2.25 & 0.65 & 2.39 & 0.16 & \textcolor{black}{2.47} & \textcolor{black}{0.25} \\
\end{tabular}
\end{table}

In order to test the applicability of our ChIMES/DFTB models to different of Si phases, we have computed the relative energies and nearest neighbor distances for several polymorphs \textcolor{black}{(Table~\ref{tab:si_polymorphs}). This includes the solid phases in our training set as well as the bc8 and the $\beta$-Sn phases for additional validation. Both pbc/ChIMES and siband/ChIMES show accurate performance for all training set phases, where the energies relative to the diamond ground-state tends to agree with DFT within 0.01~eV, and the subsequent nearest neighbor distances agree within $0.01-0.02$ \AA. The graphene phase is a small exception, where pbc/ChIMES yielded a relative energy of 0.70~eV/atom and siband/ChIMES a relative energy of 0.59~eV, compared to a value of 0.65~eV for DFT. Both pbc/ChIMES and siband/ChIMES yield an energetic minimum structure and relative energy for the bc8 phase that compares well with DFT. However, pbc/ChIMES yields a nearest neighbor distance for the $\beta$-Sn phase of 2.63 \AA\ compared to a value of 2.47 \AA\ from DFT, though the relative energy matches well. In contrast, siband/ChIMES yields a close match to the DFT results for both the nearest neighbor distance and relative energy, with values of 2.47 \AA\ and 0.25~eV. }

\begin{figure}[!htp]
\begin{center}
     \begin{subfigure}[b]{0.65\textwidth}
         \includegraphics[width=\textwidth]{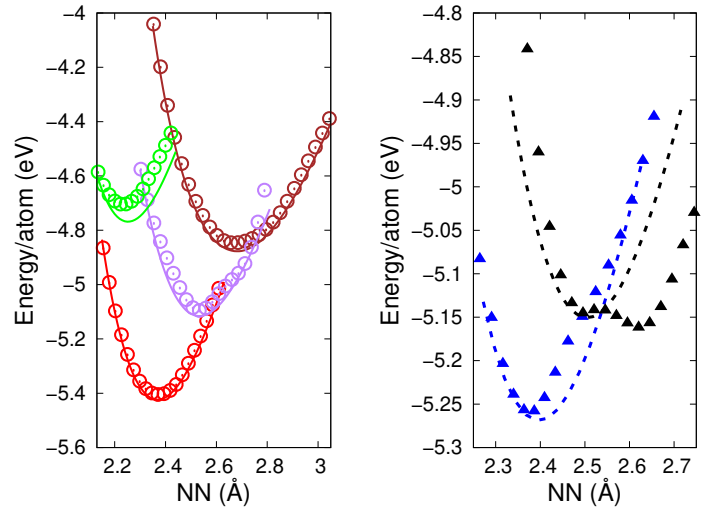}
         \caption{pbc/ChIMES}
     \end{subfigure}
     \begin{subfigure}[b]{0.65\textwidth}
         \includegraphics[width=\textwidth]{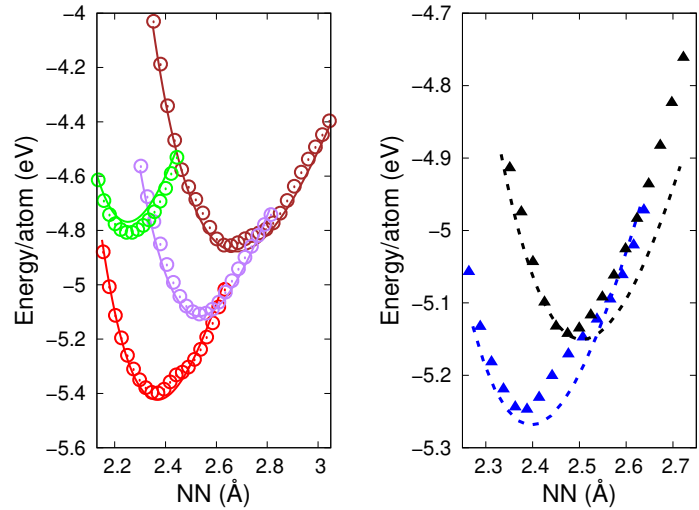}
         \caption{siband/ChIMES}
     \end{subfigure}
\caption{\label{fig:si_polymorphs} \textcolor{black}{Cold curves for several silicon polymorphs from pbc/ChIMES and siband/ChIMES DFTB models (points) compared to results from DFT (curves). The solid curves and open circles in the left panels correspond to silicon phases included in our training set, where black corresponds to the diamond phase, purple to simple cubic, brown to bcc, and green to graphene. The dashed curves and solid triangles on the right correspond to the two phases not included in our training set, where blue corresponds to the bc8 phase and black to $\beta$-Sn.}}
\end{center}
\end{figure}

Similar to previous efforts\cite{Kullgren_CCS_2021,Niehaus_NN_Si_2022}, we have determined cold energy curves under isotropic compression and expansion for all phases in this study (Fig.~\ref{fig:si_polymorphs}).  Overall, both pbc/ChIMES and siband/ChIMES yield close agreement with DFT \textcolor{black}{for phases included in our training set}. Both models have particularly close agreement for the diamond and simple cubic phases. \textcolor{black}{The siband/ChIMES model exhibits a small oscillation in the diamond and bcc cold curves at expanded volumes, which is not present in the pbc/ChIMES result. These oscillations persisted even when increasing the LASSO regularization parameter by three orders of magnitude (i.e., up to a value of 1.0).} The largest disagreement for pbc/ChIMES is with graphene, where it yields a more positive curvature at expanded densities, whereas siband/ChIMES yields closer agreement to DFT overall. Both models predict very similar agreement for the bc8 phase, where each yielded a small oscillation in the cold curve around 2.5 \AA. This is likely due to insufficient sampling of these Si-Si distances and bonding environments in our training set. \textcolor{black}{The pbc/ChIMES result for the $\beta$-Sn phase shows reasonable agreement with DFT for nearest neighbor distances below $\sim$2.5 \AA, though it yields a spurious double-well energy curve, with minima existing at both 2.47 \AA\ and 2.62 \AA. In contrast, siband/ChIMES shows much closer agreement with DFT, with the $\beta$-Sn cold curve showing some disagreement for the volume expanded states. Overall, these results indicate strong agreement for energy vs. volume relationships, which could indicate accurate force prediction from each model for the diamond phase in particular.}

\begin{figure}[!htp]
\begin{center}
\includegraphics[scale=0.4]{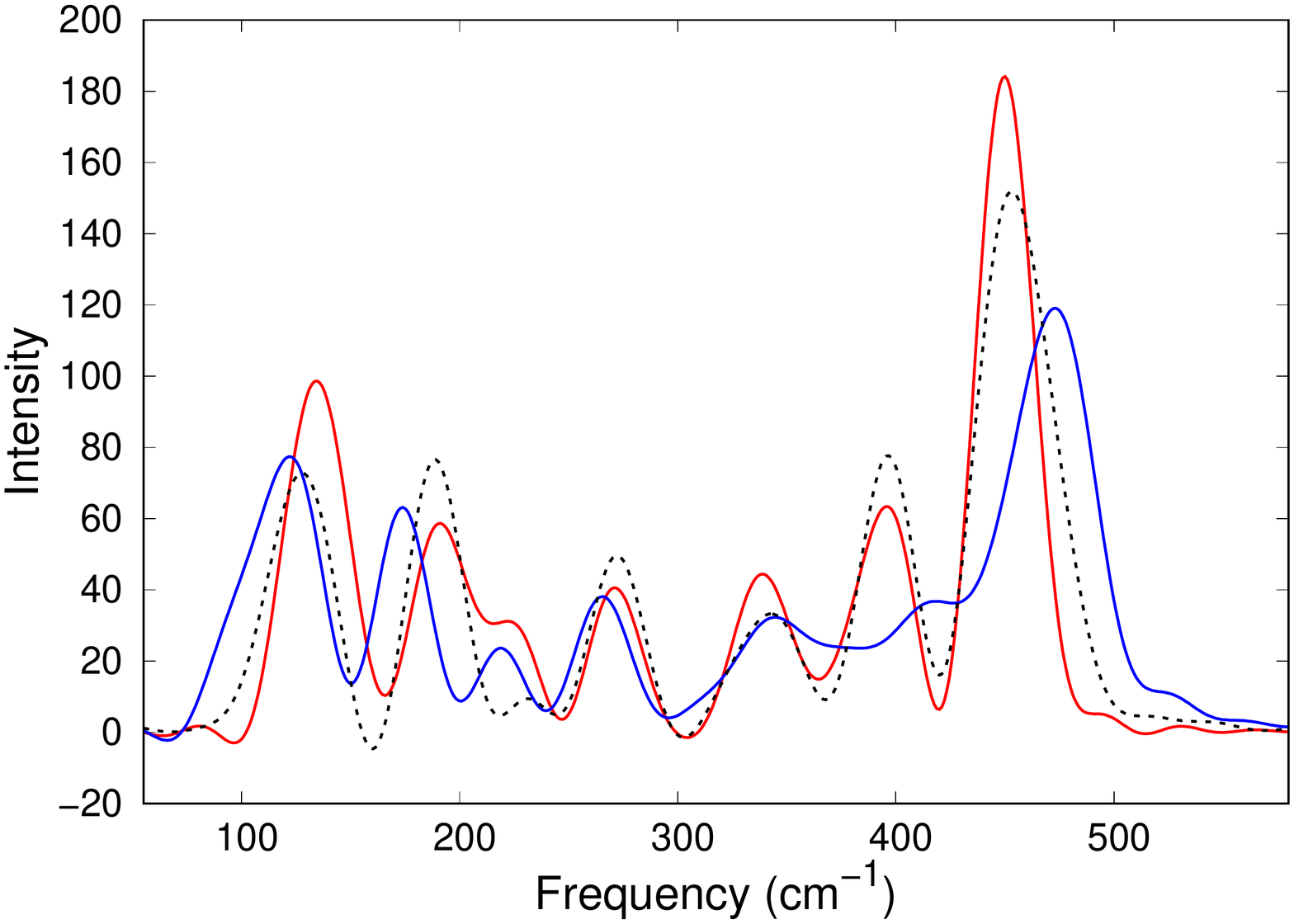}
\caption{\label{fig:si_vdos} Vibrational density of states for the Si diamond phase, computed at 600~K. The red line corresponds to pbc/ChIMES. the blue line to siband/ChIMES, and the black dashed line to DFT.}
\end{center}
\end{figure}

We now assess the force output from each model through comparison of the resulting vibrational density of states (VDOS) for the diamond phase to results from DFT (Fig.~\ref{fig:si_vdos}). These were computed from Fourier Transform of the velocity autocorrelation function which was determined from MD simulations at constant volume-temperature (NVT), conducted at 600~K, using a Nos\'e-Hoover thermostatted chain\cite{nose-1984,hoover-1985,nose_hoover_chains} and run for 15--20 ps using a timestep of 1~ps. \textcolor{black}{We note that difference between our results and previous work\cite{Niehaus_NN_Si_2022} is likely due to our determination of the VDOS from molecular dynamics simulation at elevated temperature.} Our results for pbc/ChIMES indicate fairly close agreement with DFT. Prediction of the lowest lying vibrational peak is off by only $\sim$7~cm$^{-1}$, with a value of 134~cm$^{-1}$ compared to a value of 127~cm$^{-1}$ from DFT. DFT yields a small peak at 231~cm$^{-1}$ which appears as a broad, higher intensity shoulder at 224~cm$^{-1}$ in the pbc/ChIMES spectrum. The remaining peaks in the spectrum show similarly strong agreement with some variation in the intensity of the peaks, including accurate prediction from pbc/ChIMES of the vibron peak at 450~cm$^{-1}$ compared to a frequency of 453~cm$^{-1}$ from DFT. 

In contrast, siband/ChIMES shows slightly less accurate agreement with DFT overall. The agreement for the lowest vibrational peak is fairly close, with a frequency of 120~cm$^{-1}$. The remainder of the siband/ChIMES spectrum yields an accurate overall shape of the VDOS, though with some errors in peak positions and intensities. There is some deviation in the siband/ChIMES spectrum for next two vibrational peaks, where we observe a frequency of 173~cm$^{-1}$ for the second lowest frequency peak compared to a value of 188~cm$^{-1}$ from DFT and a frequency of 217~cm$^{-1}$ for the low intensity peak after that compared to the previously mentioned DFT peak at 231~cm$^{-1}$. The siband/ChIMES spectrum yields a close match in intensity and frequency with DFT for the VDOS peak at 344~cm$^{-1}$. However, the subsequent two peaks are red shifted in frequency and lower in intensity, with values of peak positions of 413 and 472~cm$^{-1}$, compared to values of 396 and 453~cm$^{-1}$ from DFT. The improved VDOS determination from pbc/ChIMES could be due in part to its parameterization with density superposition, which has been shown to yield more accurate predictions over potential superposition\cite{Goldman_TiH2}. We note that these peak position differences discussed here correspond to small changes in energy, where 20~cm$^{-1}$ corresponds to $\sim2.5 \times 10^{-3}$~eV. Hence, it is possible that siband/ChIMES will still yield sufficiently accurate forces for some applications.

\begin{figure}[!htp]
\begin{center}
     \begin{subfigure}[b]{0.2\textwidth}
         \includegraphics[width=\textwidth]{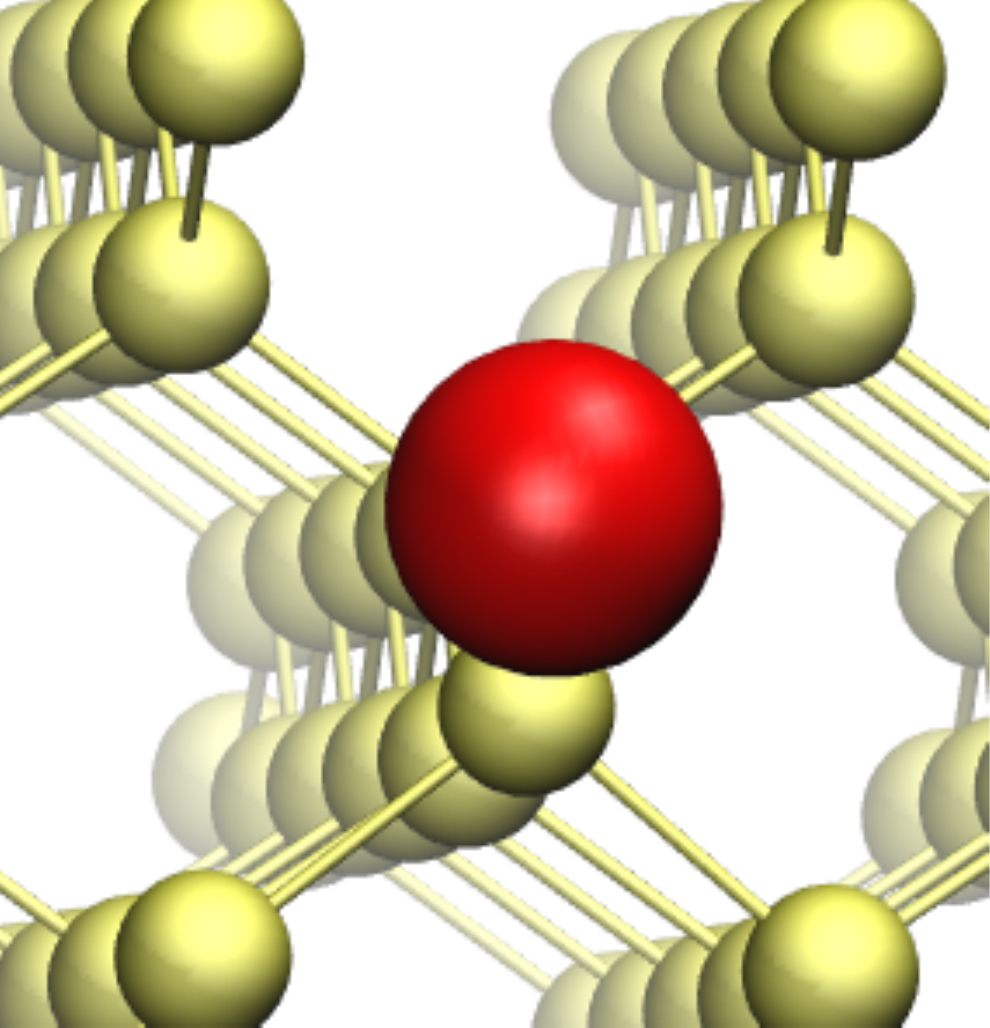}
         \caption{Vacancy}
     \end{subfigure}
     \begin{subfigure}[b]{0.2\textwidth}
         \includegraphics[width=\textwidth]{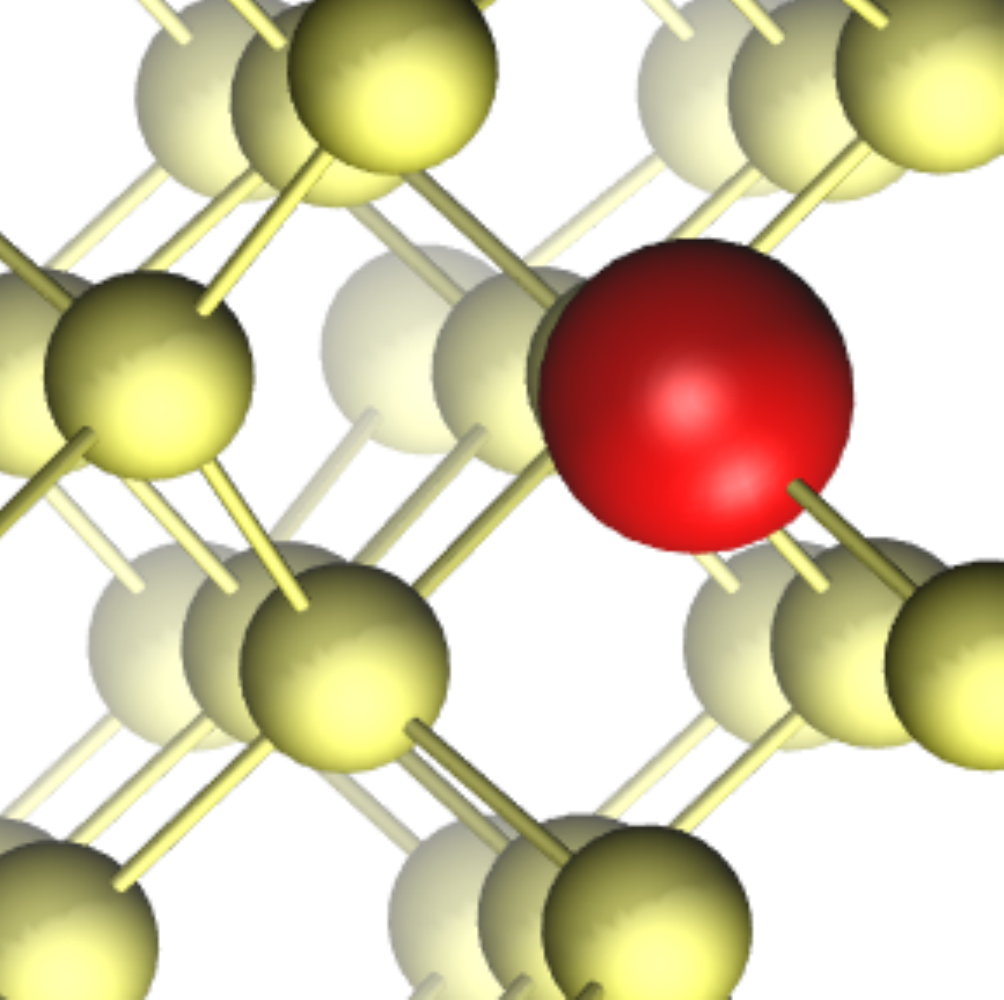}
         \caption{Tetrahedral}
     \end{subfigure}
     \begin{subfigure}[b]{0.2\textwidth}
         \includegraphics[width=\textwidth, trim={0 0 0 1cm},clip]{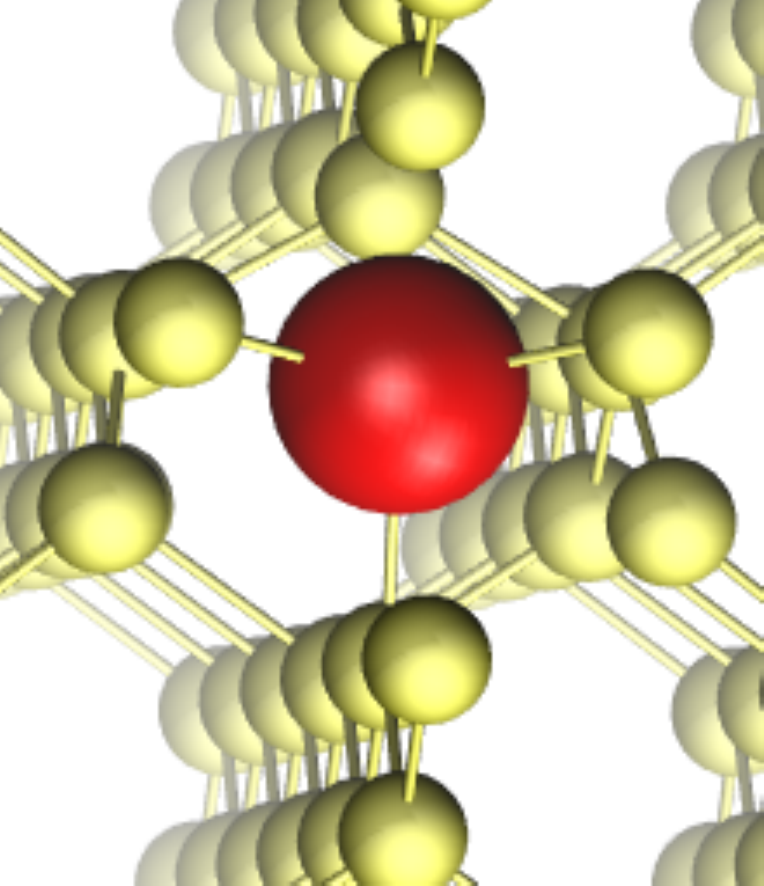}
         \caption{Hexagonal}
     \end{subfigure}
\caption{\label{fig:si_defects} Images of the diamond phase point defects investigated in this study. All defects are shown as a red sphere for the sake of clarity.}
\end{center}
\end{figure}

\begin{table}[!htp]
  \footnotesize
  \caption{Defect formation energies for the Si diamond phase. All energies are in eV.}
  \label{tab:si_defects}
\begin{tabular}{cccc}
Defect & pbc/ChIMES & siband/ChIMES & DFT (PBE) \\
\hline
\hline
vacancy & 3.45 & 4.60 & 3.84 \\
tetrahedral & 5.11 & 4.88 & 3.84  \\
hexagonal & 5.87 & 4.79 & 3.61  \\
\end{tabular}
\end{table}

Finally, we have computed defect formation energies from our DFTB/ChIMES models (Fig.~\ref{fig:si_defects}). Here, we have investigated a single Si atom vacancy as well as an interstitial atom in either a hexagonal or tetrahedral site, which were determined from use of the pymatgen software suite\cite{pymatgen_code}. The tetrahedral interstitial site occurs where an additional Si atom is coordinated by four atoms from the lattice, whereas the hexagonal interstitial site occurs when the additional Si atom resides in a hexagonal opening within the lattice. The defect formation energy $E_\mathrm{form}$ is computed as $E_\mathrm{form} = E_\mathrm{def} - N_\mathrm{def} E_\mathrm{diam}$, where $E_\mathrm{def}$ is the total energy of the defect containing system, $N_\mathrm{def}$ is the number of Si atoms in that configuration, and $E_\mathrm{diam}$ is the energy per atom of the perfect diamond phase. Similar to previous Si DFTB efforts\cite{Niehaus_NN_Si_2022}, calculations were initialized from an optimized 216 atom supercell where we retained a Monkhorst-Pack mesh of $2 \times 2 \times 2$, after which we created the point defect and optimized the ionic positions of each configuration using the same k-point mesh.
Our results indicate some agreement with previous PBE-DFT calculations from Ref.~\citenum{Niehaus_NN_Si_2022}. The pbc/ChIMES model agrees with the DFT vacancy energy within 0.4~eV, but yields results that are 1--2 eV too high for both interstitial energies. In particular, the three defect energies from pbc/CHIMES differ over a range of over 2.4~eV, with the both interstitial energies yielding larger results than that of the vacancy. In contrast, the result from DFT all lie relatively close together (within a range of 0.23~eV) and DFT exhibits equal formation energy values for the vacancy and tetrahedral interstitial. It is likely that the interstitial energies would be decreased with full accounting of d-orbital off-site interactions which are absent in the original pbc-0-3 parameter set.  The siband/ChIMES model yields defect formation energies that are consistently $\sim$1~eV too high relative to DFT. However, the siband/ChIMES results differ over an energy range of 0.28~eV, yielding improved agreement with DFT in this respect. It is likely that there would be some variation in DFT results depending on the choice of exchange-correlation function and possible inclusion of a dispersion energy correction.

Overall, our we able to create two new DFTB/ChIMES models that more closely approach a single-purpose approach for silicon phases under different conditions. The pbc/ChIMES model appears to yield a somewhat improved description of atomic forces \textcolor{black}{though the description of the $\beta$-Sn phase could show improvement with targeted training data. The siband/ChIMES model yields an overall accurate description of the various silicon phases discussed here, as well as} a more systematically consistent defect formation energies that could make it preferable for some calculations. As mentioned, some of the limitations of the pbc/ChIMES model could possibly be overcome through inclusion of d-orbital two-center interactions in the corresponding Slater-Koster file. Regardless, we now provide a repulsive energy for the siband-1-1 parameter set, which will allow its use for structural relaxations and/or dynamics calculations in addition to its accuracy for electronic properties. It is possible that the slightly longer cutoff radius for our ChIMES $E_\mathrm{Rep}$ could be mitigated through optimization of the choice of DFTB confining radii (discussed in the next section). Future improvement of these models could also involve inclusion of data from MD simulations of amorphous or defect containing systems at different temperatures and pressures.

\subsection{Semi-automated Workflow for DFTB/ChIMES Model Creation}

\begin{figure}[!htp]
\begin{center}
\includegraphics[scale=0.5]{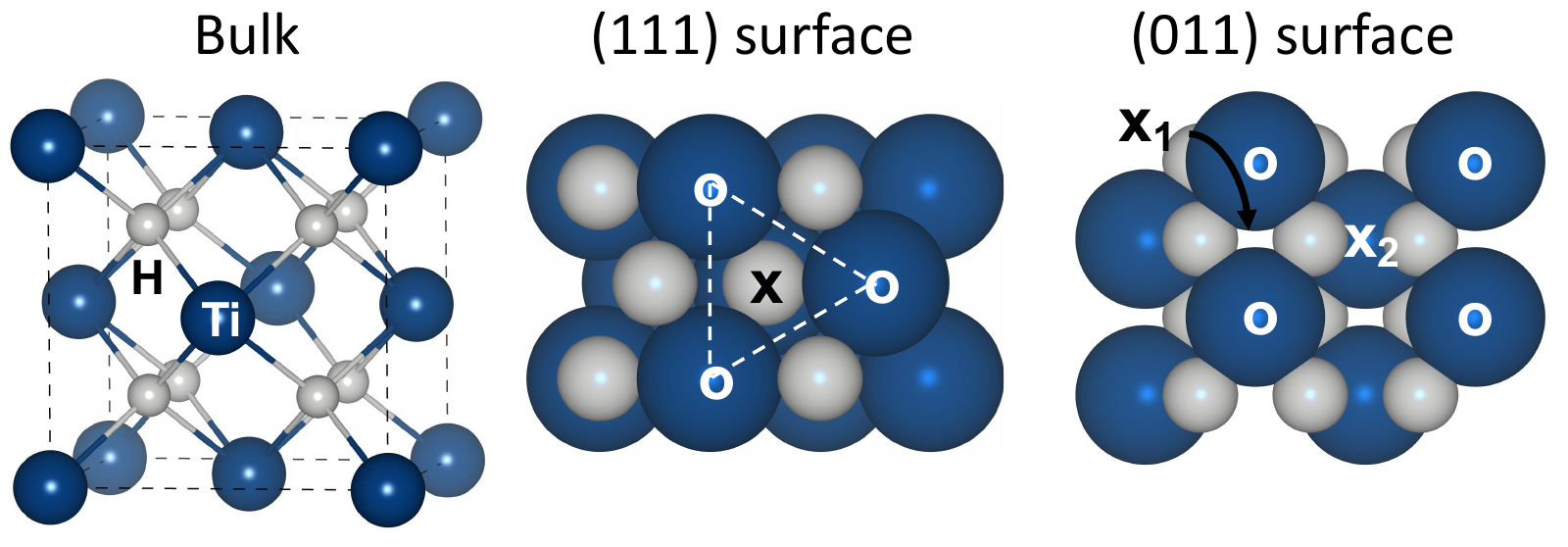}
\caption{\label{fig:tih2}  Snapshots of several  TiH$_2$ configurations discussed in this study. Relevant surface adsorption sites include the Top (`O') and Hollow (`X') sites for the (111) facet and Top, Bridge-1 (`$\mathrm{X}_1$') and Bridge-2 (`$\mathrm{X}_2$') sites for the (011) facet. Reprinted with permission from \textit{Journal of Chemical Theory and Computation} \textbf{2021} \textit{17} (7), 4435-4448. Copyright 2021, American Chemical Society.}
\end{center}
\end{figure}

In this subsection, we summarize previous work on TiH$_2$\cite{Goldman_TiH2} which indicates the utility in using a ChIMES $E_\mathrm{Rep}$ in a semi-automated fashion to screen for optimal confining radii in a Slater-Koster file parameterization. TiH$_2$ has a number of industrial uses as a functional material, including in hydrogen storage alloys\cite{Kitabayashi_MgH2-TiH2}, superconductors\cite{Shanavas_TiH2_2016}, and as a blending agent for porous foams\cite{Peng_TiH2_blends_2016}. Its ground-state structure exhibits face-centered-cubic (fcc) symmetry, with the (111) facet computed to have the lowest surface energy (Fig.~\ref{fig:tih2}). Several adsorption sites are illustrated, including Top (directly above a Ti atom), Hollow (in an interstitial cavity), and several Bridge sites (existing in between Ti-Ti and H-H nearest neighbors) sites. TiH$_2$ is a somewhat ideal system for DFTB model development in that DFT calculations on small  supercells are relatively tractable, which allows for straightforward validation. DFT calculations though are generally too computationally inefficient for the larger supercells needed to model grain boundaries and crystalline defects at sufficiently low concentration, allowing for several applications of a new TiH$_2$ DFTB model in future studies.

\begin{figure}[!htp]
\begin{center}
\includegraphics[scale=0.5]{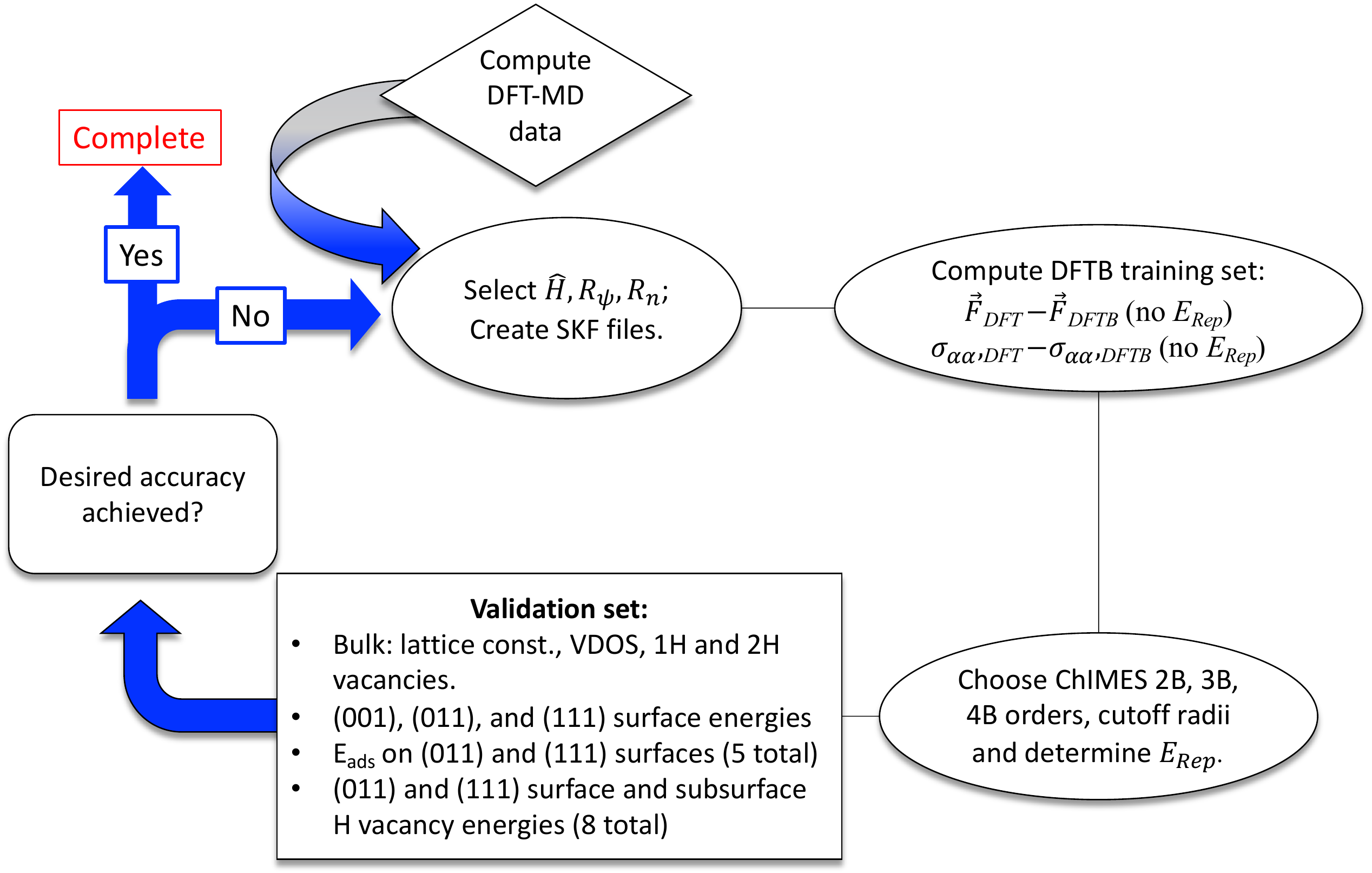}
\caption{\label{fig:flowchart} Flowchart for creation of DFTB $E_\mathrm{rep}$ models through ChIMES force field parameterization. Reprinted with permission from \textit{Journal of Chemical Theory and Computation} \textbf{2021} \textit{17} (7), 4435-4448. Copyright 2021, American Chemical Society.}
\end{center}
\end{figure}

Here, we have leveraged rapid ChIMES $E_\mathrm{Rep}$ optimzation by creating a workflow that allowed us to perform a semi-exhaustive search of all DFTB and ChIMES hyperparameters (Fig.~\ref{fig:flowchart}). We first compute a matrix of thirty Slater-Koster files from titanium wavefunction confining radii ($R^{Ti}_\psi$) and density confining radii ($R^{Ti}_n$) sampled over a range of $3.2 \leq R^{Ti}_\psi \leq 5.0$ au and $6.0 \leq R^{Ti}_n \leq17.0$ au.  Hydrogen interaction parameters were taken from the miomod-hh-0-1 parameter set. Model down selection could then be performed over the entire grid Slater-Koster tables through comparison to our selected validation data, which allowed us to determine optimal ChIMES polynomial orders and the LASSO regularization parameter.

For this work, our DFT training set consisted of MD simulations of the TiH$_2$ unit cell (i.e., 4 formula units) run in the canonical ($NVT$) ensemble at 400~K and initial pressures up to $100$~GPa. DFT calculations were performed with VASP, using the PBE functional. We also include several unit cell simulations with a single hydrogen vacancy or interstitial site in order to sample different Ti-H bonding environments. ChIMES $E_\mathrm{Rep}$ optimization was then performed on the atomic forces and the diagonal of the stress tensor. Our training set contained a total of 153 MD configurations, which comprised only 5988 data points. Our validation set consisted of the bulk lattice constant, single and double hydrogen vacancy energies, and vibrational density of states, as well as surface energies for the (001), (011), and (111) facets, hydrogen surface adsorption energies, and surface and subsurface hydrogen vacancy energies. Further details of all aspects of our calculations are discussed in Ref.~\citenum{Goldman_TiH2}.  

DFTB$+$ calculations were performed using self-consistent charges (SCC)\cite{DFTB+_scc}. All minimum and cutoff radii for the ChIMES $E_\mathrm{Rep}$ were set to include the first coordination shell sampled in our training set, only: $2.5 \leq r_{TiTi} \leq 3.5$ $\,$\AA$\,$ and $1.5 \leq r_{HTi}  \leq 2.5$ $\,$\AA$\,$. We use values of $\lambda_{TiTi} = 3.0\,$\AA$\,$ and $\lambda_{HTi} = 2.0\,$\AA$\,$ for the Morse-like coordinate transforms. H-H repulsive interaction were not sampled in our training set and were thus also taken from the miomod-hh-0-1 parameter set. \textcolor{black}{ChIMES $E_\mathrm{rep}$ determination for each of the thirty Slater-Koster parameterizations for TiH$_2$ required less than one minute on a single Intel Xeon E5-2695 v4 processor. The rapidity of the optimization step greatly facilitated down selection of DFTB Hamiltonian parameters.}

\begin{figure}[!htp]
\begin{center}     
     \begin{subfigure}[b]{0.65\textwidth}
         \includegraphics[width=\textwidth]{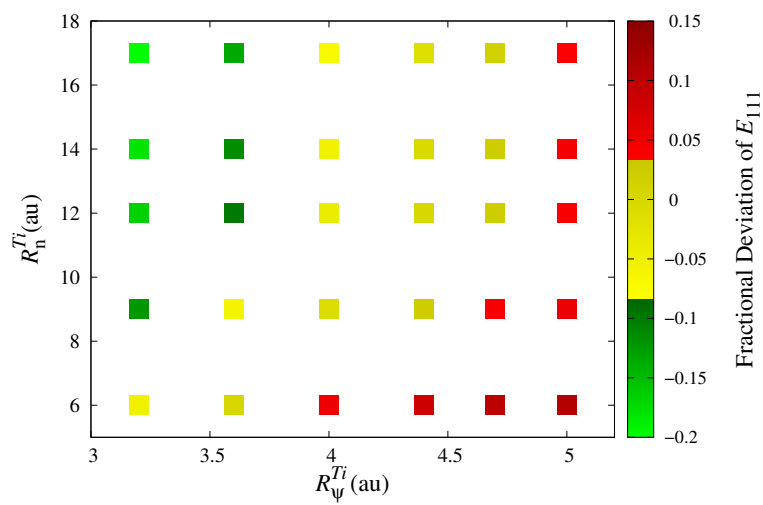}
         \caption{}
     \end{subfigure}
     \begin{subfigure}[b]{0.65\textwidth}
         \includegraphics[width=\textwidth]{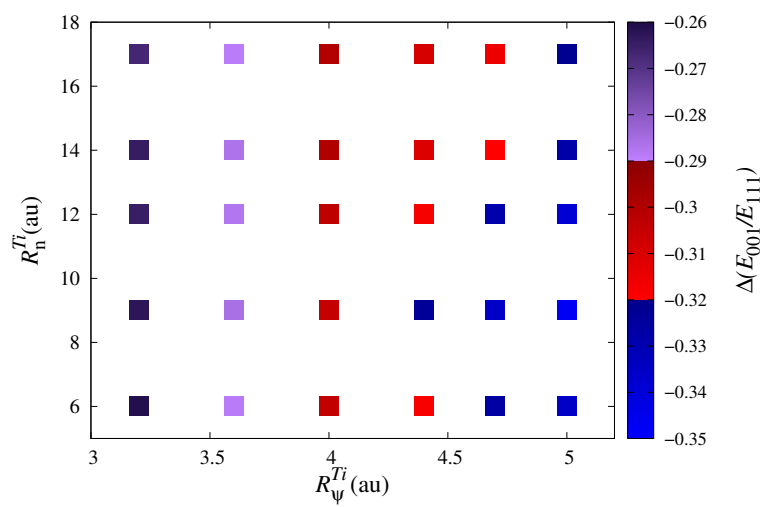}
         \caption{}
     \end{subfigure}
\caption{\label{fig:radius_sweep} Results for sweep of values of $R^{Ti}_{\psi}$ and $R^{Ti}_{n}$, where the ChIMES $E_\mathrm{Rep}$ was determined with a 2B order of 12 and 3B order of 8. The top panel (a) corresponds to the fractional deviation of the surface energy, $\left(E^{DFTB}_{111} - E^{DFT}_{111}\right)/E^{DFT}_{111}$, and the bottom panel (b) to the deviation of $\left(E_{001}/E_{111}\right)$ relative to DFT. Reprinted with permission from \textit{Journal of Chemical Theory and Computation} \textbf{2021} \textit{17} (7), 4435-4448. Copyright 2021, American Chemical Society.}
\end{center}
\end{figure}

Our results for a subset of our validation data (Fig.~\ref{fig:radius_sweep}) allow us to describe general trends regarding the confining radii. We observe an approximate linear relationship between $R^{Ti}_\psi$ and $R^{Ti}_n$ in terms of the accuracy of the $E_{111}$ energy, where the most accurate surface energy results from either small or large choice for both radii. All of the DFTB/ChIMES models created in this iteration under-predict the $\left(E_{001}/E_{111}\right)$ ratio (i.e., the ratio of highest to lowest surface energies in our study) relative to our DFT calculations, where we observe values of 1.35--1.44 compared to the DFT ratio of 1.70. We note that is is likely in part due to the surface dipole moment present in our construction of the (001) facet.  

Our final set of hyper-parameter values includes \{$R^{Ti}_\psi=3.6\, \mathrm{au}$, $R^{Ti}_n=6.0\, \mathrm{au}$\} and \{${{\cal O}_{\rm{2B}}} = 8$, ${{\cal O}_{\rm{3B}}} = 4$\} and LASSO $\alpha$ of 10$^{-3}$, with hydrogen RMS force errors of 0.076~eV/\AA$\,$ titanium errors of 0.056~eV/\AA$\,$, and an error value of 0.35~GPa for the stress tensor diagonal. Our optimum DFTB/ChIMES model yields an accurate lattice constant, with an error of $\sim$0.4\% relative to the DFT result. However, DFTB/ChIMES yields a hydrogen bulk vacancy energies that are systematically $\sim$0.5~eV too small. This was true for both bulk single and di-vacancies, as well as surface and subsurface values. It is possible that this could be corrected in subsequent work through either an expanded training set or inclusion of a larger basis set or higher-bodied terms in the Hamiltonian.\cite{Goldman15-CPL}.

Overall, DFTB/ChIMES yields accurate surface energies for the surface energies investigated in this study (Table~\ref{tab:surface_val}), with nearly identical results to DFT for the (011) and (111) facets. As mentioned, the deviation in the (001) surface energy could be due in part to the internal electric field on the (001) surface configuration studied here, where DFTB can underestimate surface electrostatic interactions\cite{Witek_DFTB_charges}. Our DFTB/ChIMES results show similarly strong agreement with hydrogen surface adsorption energies (Table~\ref{tab:surface_Hads}), with overall favorable agreement with DFT for all sites studied here.

\begin{table}[!htp]
  \footnotesize
  \caption{TiH$_2$ surface energies (in eV/\AA$^2$). Reprinted with permission from \textit{Journal of Chemical Theory and Computation} \textbf{2021} \textit{17} (7), 4435-4448. Copyright 2021, American Chemical Society.}
  \label{tab:surface_val}
\begin{tabular}{ccc}
Surface & DFTB/ChIMES & DFT \\
\hline
\hline
111& 0.080 & 0.080 \\
011& 0.105 & 0.101 \\
001& 0.114 &  0.136 \\
\end{tabular}
\end{table}

\begin{table}[!htp]
  \footnotesize
  \caption{Surface hydrogen adsorption energies on TiH$_2$ surface sites (in eV). Reprinted with permission from \textit{Journal of Chemical Theory and Computation} \textbf{2021} \textit{17} (7), 4435-4448. Copyright 2021, American Chemical Society.}
  \label{tab:surface_Hads}
\begin{tabular}{cccc}
Surface & Site & DFTB/ChIMES & DFT \\
\hline
\hline
111& Top & -1.888 & -1.760 \\
& Hollow & -2.081 & -2.440  \\
011& Top & -2.383 & -2.332  \\
& Bridge-1 & -2.154 & -2.442 \\
& Bridge-2 & -2.132 & -2.342 \\
\end{tabular}
\end{table}

Our results indicate DFTB/ChIMES has relatively small data requirements and can allow for semi-exhaustive optimization of the confining radii, which is traditionally a difficult task for DFTB model development. Our approach can be applied to complex systems where the underlying crystal symmetry can be broken due to the presence of either point defects or surfaces. Our current DFTB/ChIMES model also yields accurate results for bulk $\alpha$-Ti and gas phase TiH$_4$ (not shown here), indicating a possibly high degree of transferability. In addition, the small training set requirements for ChIMES optimization could provide significant advantages for systems requiring a high degree of computational effort, such as high-Z and/or magnetic materials, where DFT data for training can be exceedingly difficult to generate. 

\subsection{$\Delta$-learning to Enhance the Accuracy of DFTB for Organic Materials}

In this subsection we review our recent efforts to leverage a high-level quantum chemical database to create an ``out-of-the-box'' model with accuracy beyond standard DFT approaches (e.g., PBE) that is generally applicable to many organic molecular systems\cite{Pham_DFTB_JPCL}. In this work, we have used the ANI-1x quantum chemical data set\cite{ANI-1ccx,ANI-dataset} to create a DFTB/ChIMES model that approaches hybrid-functional and/or coupled cluster accuracy. Here, ChIMES is used as a $\Delta$-learning potential where we have included it as an extra energy term to the 3ob-3-1 parameterization\cite{DFTB3,Gaus12}, which includes third-order charge fluctuation terms in the DFTB energy. This parameterization is known to yield reliable accuracy for many organic molecules and thus was a reasonable starting point for our efforts. We have found that the advantages of ChIMES over a neural network approach are two-fold: (1) the training set requirements of ChIMES are significantly lower, where only a small fraction of the ANI-1x dataset was required to achieve a high degree of accuracy, and (2) our ChIMES potential requires two orders of magnitude fewer parameters than several recent NN-based semi-empirical \textcolor{black}{approaches\cite{ISO34_DFTB_DTNN,Tretiak_PNAS_2022}. As mentioned, ChIMES can yield advantages in some situations in that it is systematically improvable due to its underlying Chebyshev polynomial basis set and that is amenable to regularization methods not available to neural networks due to its linear parameterization.} 

The original ANI-1x database was developed for the creation of ML-based general-purpose organic potentials where the data set was determined
through an active learning process\cite{ANI-dataset}, resulting in approximately 5 million molecular equilibrium and non-equilibrium configurations. Our $\Delta$-learning optimization used an iterative approach by first creating a subset of ANI-1x called ``sub\_ANI-1x'' that only contained results computed
from CCSD(T) (coupled-cluster considering single, double, and perturbative triple excitations) and using a well-known hybrid functional, $w$B97X\cite{wB97X}. This corresponded to 459,464 molecular confirmations
from computed from 1895 unique molecules, or $\sim$10\% of the original ANI-1x database. We note that there are no atomic force data from
CCSD(T)/CBS calculations. Hence, we used $w$B97X results computed with a large basis set (def2-TZVPP) data for fitting purposes, with the remainder of the data set available for
validation.

\begin{figure}[!ht]
\centering
\includegraphics[width=.85\textwidth]{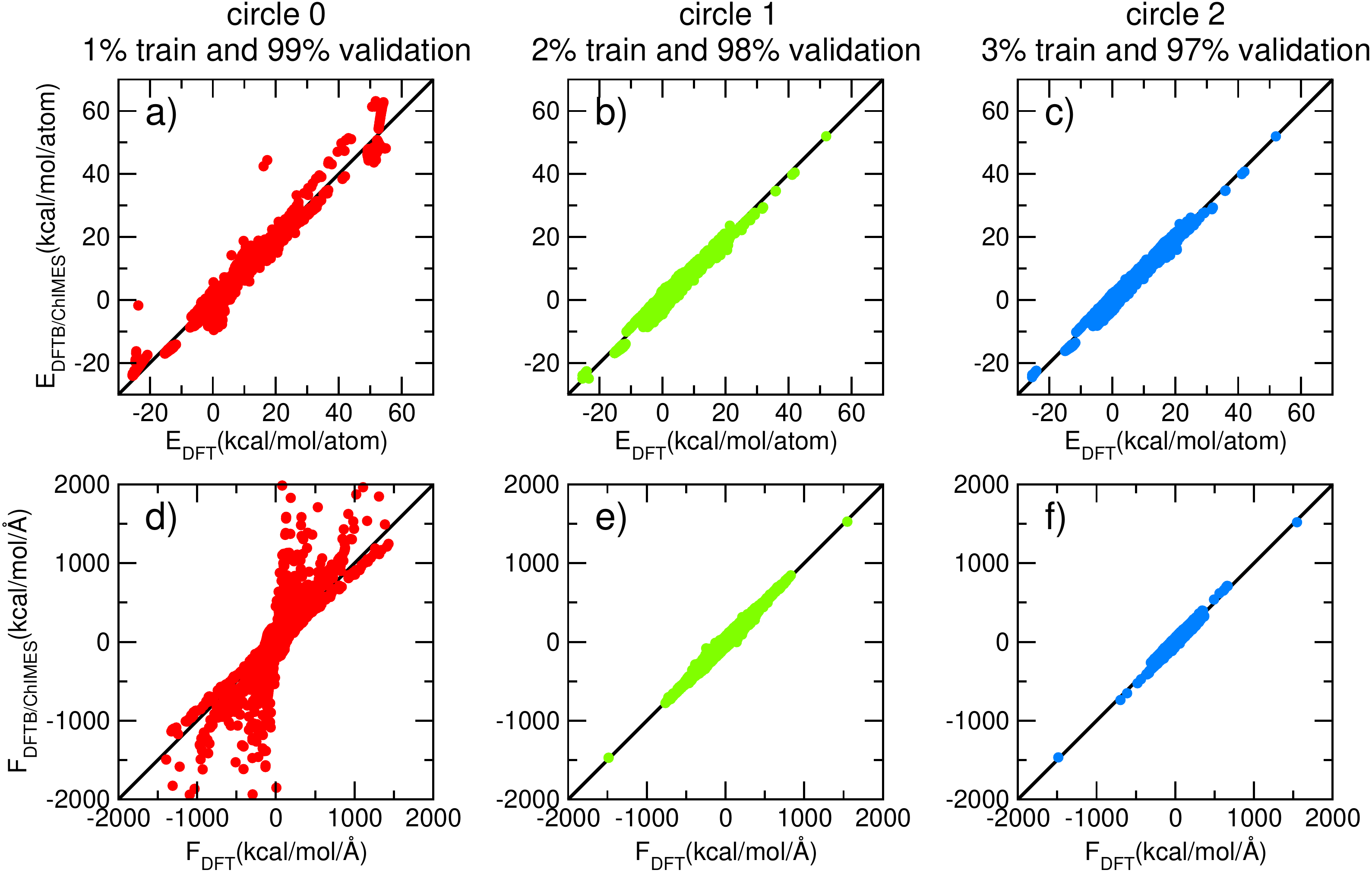}
\caption{
Comparison of energies per atom (panels a, b, c) and forces (panels d, e, f) predicted by DFT ($w$B97X) and DFTB/ChIMES for all configurations in the validation set. The dataset used here is `sub\_ANI-1x', $\sim$10\% of the full ANI-1x. Reprinted with permission from \textit{J. Phys. Chem. Lett.} \textbf{2022} \textit{13} (13), 2934-2942. Copyright 2022, American Chemical Society.}
\label{fig1_validation}
\end{figure}

We then used an iterative approach to ChIMES optimization (Fig.~\ref{fig1_validation}) where we first randomly selected only 1\% of
sub\_ANI-1x and performed an initial ChIMES optimization. Validation calculations agains the remainder of sub\_ANI-1x resulted in some large
deviations in the computed energies and forces. We then selected an additional equivalent of 1\% of the data set from configurations with the
highest force deviations and added them to our training set and repeated the process, where each increment of the training process would include the equivalent of
an additional 1\% of sub\_ANI-1x. Our DFTB/ChIMES $\Delta$-learning was converged after three iterations of our optimization scheme, using only 3\% of sub\_ANI-1x or 0.3\% of the original ANI-1x database. Our model was ultimately validated against the entire sub\_ANI-1x data set, though its size is
somewhat arbitrary and it is possible that a smaller subset of ANI-1x could have been used for our purposes.

Our final model used ChIMES polynomial orders of \{2B = 24, 3B = 10, 4B = 0\} with a somewhat long radial cutoff of 4.0 \AA\ used for all atom pairs. This longer cutoff helped account for some dispersion interactions that would otherwise be absent from standard DFTB calculations, though future efforts will involve shorter cutoffs combined with a dispersion interaction model. Further details about our ChIMES model for organics can be found in the Supporting Information in Ref.~\citenum{Pham_DFTB_JPCL}. Ultimately, our DFTB/ChIMES model resulted in 5546 parameters and was trained to $\sim$372k data points. This is in contrast to the recently developed AIQM1 semi-empirical quantum model, which utilizes an NN trained to the entire ANI-1x data set, resulting in 322,660 parameters. Similarly, a recent DFTB-NN approach using deep-tensor neural networks used a training set of $\sim$800k data points, resulting in 228,865 parameters.

\begin{table*}[!ht]
\caption{
Performance of DFTB and DFTB/ChIMES in predicting reference energies and/or atomic forces in the GDB-10to13, ISO34, and GDML data set. The MAE and RMSE for the energies and forces (labeled with subscripts `E' and `F') are in kcal/mol and kcal/mol-\AA, respectively. Reference molecular energies and atomic forces in the GDB-10to13 data set are at the $w$B97X/6-31G* level of theory. Isomerization energies in the ISO34 data set are a mixture of experimental- and CCSD(T) extrapolation energies. The CCSD(T)/cc-pVTZ atomic forces of 2000 configurations of ethanol in the GDML data set are used for comparison.  Reprinted with permission from \textit{J. Phys. Chem. Lett.} \textbf{2022} \textit{13} (13), 2934-2942. Copyright 2022, American Chemical Society.
}
\begin{tabular}{ |c|cc|c|c| }
\hline
\hline
& \multicolumn{2}{|c|}{GDB-10to13} & ISO34 & GDML   \\
\hline
\hline
method      & MAE$\rm{_E}$/RMSE$\rm{_E}$ & MAE$\rm{_F}$/RMSE$\rm{_F}$ & MAE$\rm{_E}$/RMSE$\rm{_E}$ & MAE$\rm{_F}$/RMSE$\rm{_F}$\\
\hline
DFTB                       & 9.10/11.70  & 6.34/9.85 & 3.69/4.96 & 4.52/6.12 \\
DFTB/ChIMES                & 3.57/4.72   & 3.62/5.33 & 2.06/2.56 & 2.72/3.61  \\
\hline
\hline
ANI-1\cite{ANI-1x_COMP6}         & 3.12/4.74   & 3.96/7.09 & -         & - \\
ANI-1x\cite{ANI-1x_COMP6}        & 2.30/3.21   & 3.67/6.01 & -         & - \\
\hline
\hline
DFTB-NN$\rm{_{rep}}$
\cite{ISO34_DFTB_DTNN}     & -           & -         & 2.21/3.30 & - \\
PBE0\cite{ISO34_DFTB_DTNN} & -           & -         & 1.82/2.48 & - \\
\hline
\hline
\end{tabular}
\label{Table_test_sets}
\end{table*}

We then tested the transferability of our DFTB/ChIMES model through comparison to different quantum chemical data that were computed at the $w$B97X or CCSD(T) level but were not a part of ANI-1x (Table \ref{Table_test_sets}). For example, the GDB-10to13 data set \cite{ANI-1x_COMP6} consists of the molecular energies and forces at the $w$B97X level of nearly 3000 molecules containing 10-13 C, N, or O atoms for a total of 47,670 configurations.  Our DFTB/ChIMES model exhibits a 60\% reduction in the mean average error (MAE) and RMSE error in the energies and a 45 \% decrease in the forces over standard DFTB. The accuracy of DFTB/ChIMES is similar to values from the ANI-1 and ANI-1x neural network interatomic potentials\cite{ANI-1x_COMP6} (i.e., stand-alone potentials without explicit quantum mechanical elements), and are smaller than the variations between $w$B97X itself and higher levels of theory such as CCSD(T) and MP2 (4.9/5.9 kcal/mol for energies and 4.6/5.9 kcal/mol-\AA~for forces)\cite{ANI-1ccx}.

Our DFTB/ChIMES model is validated against additional CCSD(T) reference data from the ISO34 data set \cite{ISO34}, which consists of energies of 34 isomers containing the elements C, H, N, and O. We observe that the accuracy of DFTB/ChIMES is much better than that for standard DFTB, is slightly improved over that from DFTB-NN$\rm{_{rep}}$, and approaches the PBE0 data given in Ref.~\citenum{ISO34_DFTB_DTNN}. To test the performance of our model on high accuracy force data specifically, we compare DFTB/ChIMES with the CCSD(T)/cc-pVTZ data for 2000 configurations of ethanol in the GDML data set \cite{GDML} (54,000 data points total). Again our DFTB/ChIMES gives an improvement over standard DFTB as MAE and RMSE are both reduced by $\sim$40\%. A direct force comparison to DFTB-NN$\rm{_{rep}}$ or the ISO34 reference was unavailable. Additional validation of our model included calculation of the n-butane dihedral potential and correct prediction of the energetic ordering of coumarin molecular crystals.

\begin{figure}[!ht]
\centering
\includegraphics[width=.85\textwidth]{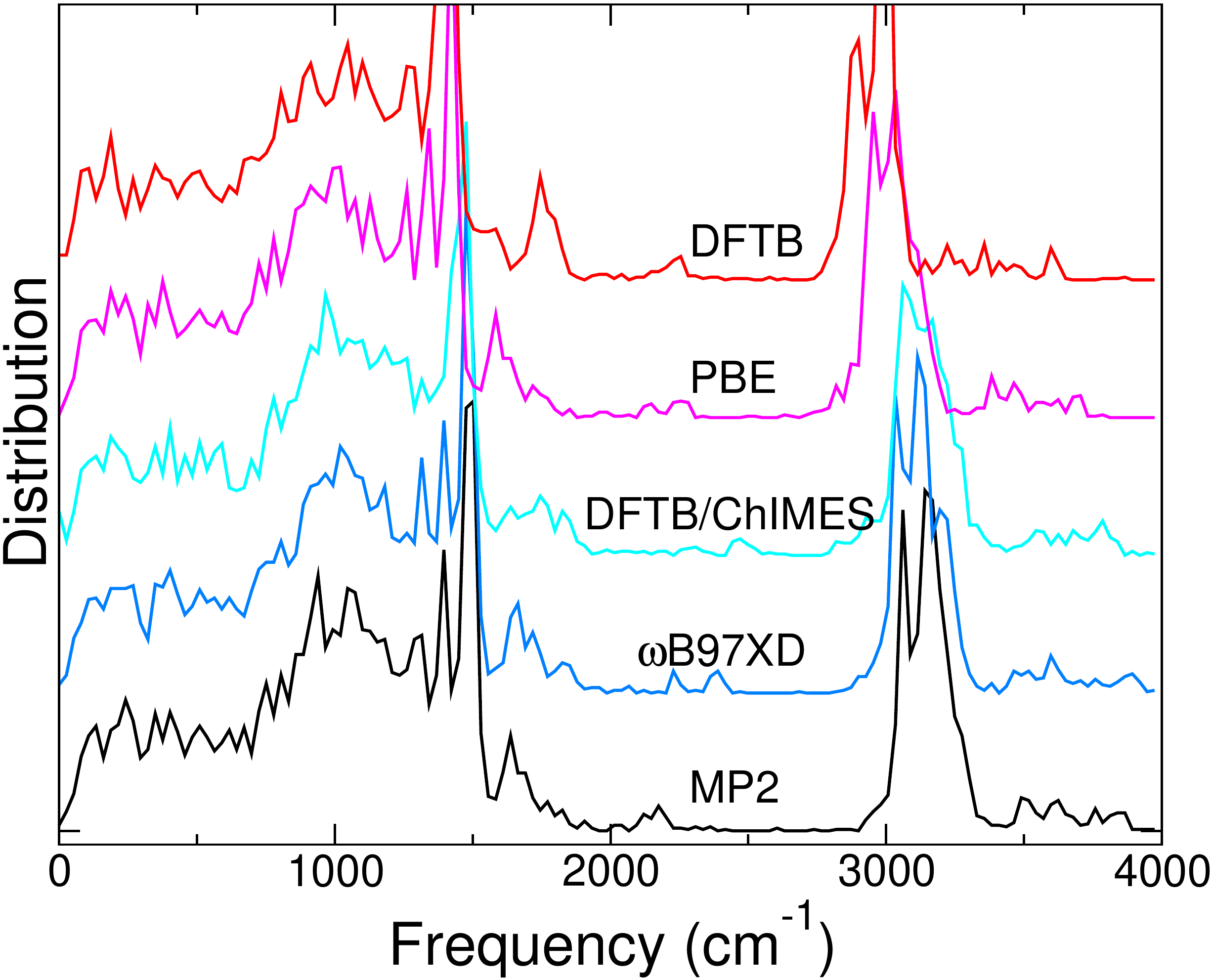}
\caption{
The distribution of the calculated frequency values using DFTB and DFTB/ChIMES for 342 neutral molecules taken from the CCCBDB database. The MP2 and DFT (PBE and $w$B97XD) calculations using cc-pVTZ basis set in the CCCBDB are selected for comparison.  Reprinted with permission from \textit{J. Phys. Chem. Lett.} \textbf{2022} \textit{13} (13), 2934-2942. Copyright 2022, American Chemical Society.
}
\label{fig3_vib_freq}
\end{figure}

We have also validated DFTB/ChIMES against vibrational frequencies of 342 gas-phase molecules from the Computational Chemistry Comparison and Benchmark Database or CCCBDB (https://cccbdb.nist.gov/), computed with MP2/cc-pVTZ and $w$B97XD (with dispersion correction), amongst other methods (Fig.~\ref{fig3_vib_freq}). Here, DFTB/ChIMES yields errors in the frequency
prediction of MAE/RMSE = 36/61 cm$^{-1}$, indicating improved accuracy over PBE and with similar accuracy to accuracy to $w$B97XD. In all of our validation tests, DFTB/ChIMES shows marked improvement over standard DFTB and PBE, and shows similar accuracy to results from $w$B97X or other higher-levels of theory. Further details of all validation calculations are provided in Ref.~\citenum{Pham_DFTB_JPCL}.

Lastly, though the DFTB/ChIMES model developed here is trained on gas phase molecular data, we have also tested its performance in reproducing the structural properties of bulk graphite and diamond.  We compare predicted density and lattice parameters from different methods in Table~\ref{Table_carbon_polymorphs}. For graphite, all computational models considered here give an accurate description of the in-plane lattice parameters. DFTB and PBE overestimate the interlayer separation ($c/2$) by over 25\% and 30\%, respectively, due to their under-prediction of dispersion interactions. DFTB/ChIMES predicts the lattice parameters and density in excellent agreement with the experimental value, with a deviation of less than 1\%. For diamond, the computed values using DFTB, DFTB/ChIMES, and PBE-DFT differ by $\sim$1\% from experimental values for lattice parameters and $\sim$3\% for the density. \textcolor{black}{The improved accuracy of DFTB/ChIMES for the graphite interplanar distance is due in part to the relatively long cutoff radius. Future efforts involve shorter cutoff radii and use of an explicit van der Waals interaction model.}

\begin{table*}[!ht]
\caption{
Comparison of predicted density and lattice parameters of graphite and diamond for DFTB, DFTB/ChIMES, PBE-DFT with experimental data.  Reprinted with permission from \textit{J. Phys. Chem. Lett.} \textbf{2022} \textit{13} (13), 2934-2942. Copyright 2022, American Chemical Society.}
\begin{tabular}{ llccccccccccc }
\hline
\hline
        phase & method  &  density (g/cm$^{3}$) & $a$(\AA) & $c/2$(\AA)     & \\
\hline
\hline
graphite & Expt.\cite{Zhao89}  & 2.26 & 2.462 & 3.356 \\
         & PBE-DFT\cite{Bucko10} & 1.71 & 2.470 & 4.420 \\
         & DFTB/ChIMES                    & 2.25 & 2.461 & 3.379 \\
         & DFTB                           & 1.77 & 2.474 & 4.248 \\
\hline
diamond  & Expt.\cite{Zhao89}           & 3.51 & 3.567 \\
         & PBE-DFT\cite{Goldman12} & 3.48 & 3.580 \\
         & DFTB/ChIMES                            & 3.42 & 3.600 \\
         & DFTB                                   & 3.42 & 3.600 \\
\hline
\hline
\end{tabular}
\label{Table_carbon_polymorphs}
\end{table*}

\textcolor{black}{Ultimately, we have shown that ChIMES can be used to extend DFTB to hybrid functional accuracy or greater for structural, energetic, and force prediction.} DFTB/ChIMES has the capability of reproducing vast quantities of high-level reference data while requiring only a small fraction of it for training. On the basis of the results presented here, DFTB/ChIMES represents a promising direction for developing general purpose quantum models that are applicable to a wide range of materials and conditions. The small training set required as well as the small number of potential parameters relative to neural network methods could yield significant advantages for future development of computational efficient models with up to coupled cluster accuracy. The ease of parameterization and transferability of DFTB/ChIMES allows for high-level quantum theory accuracy in systems where traditional wavefunction or hybrid functional methods are far too computationally intensive for intensive use.

\section{Discussion and Future Work}

ChIMES was initially developed as a method for creating many-body force fields for molecular dynamics simulations. However, it has also proven robust as a repulsive energy for DFTB models, where the standard two-center approach for both quantum mechanical and repulsive terms can be insufficient for many systems. The strength in ChIMES as an element of a semi-empirical quantum model  or MD model lies in its use of linear combinations of many-body Chebyshev polynomials, where the nearly optimal nature of the polynomials as well as the linear least-squares fitting allow for rapid optimizations that require far fewer parameters and significantly smaller data sets than the neural network models reviewed here. In addition, ChIMES adds very little extra computational time to DFTB calculations, where the matrix diagonalization and SCC convergence use the vast majority of the CPU effort.

Future work will involve extending ChIMES to systems with four or more elements, where development of training sets and proper validation approaches remains an open question. It is likely that these ChIMES models will require larger data sets and the potentials themselves will have significantly more parameters than those presented in this work due to the combinatorial effect of forming many-body clusters with different possible combinations of elements. Determination of $E_\mathrm{Rep}$ for these systems will likely yield significant advantages over pure interatomic potentials due to the short-ranged nature of the repulsive energy as well as the general accuracy of the quantum mechanical parts of DFTB. Both of these considerations make creation of DFTB/ChIMES model in general more tractable than optimizing ChIMES on its own as an atomistic force field. ChIMES does not take into account the charge state or spin of a molecule.  That could limit the accuracy of DFTB/ChIMES calculations for excited states or ions.  DFTB/ChIMES can serve as either a stand-alone model for running dynamics and determining physical and chemical properties of a system, or as a surrogate for DFT in a ``boot-strapping'' optimization, where it can serve to generate reasonably high fidelity training data for pure ChIMES MD models. Overall, our approach can be used to enhance the speed of quantum accurate predictions for both molecular and condensed matter systems, where there is a historic reliance on computationally intensive quantum simulations for predictions of chemical and physical properties related to experiments.

\begin{acknowledgments}

This work performed under the auspices of the U.S. Department of Energy
by Lawrence Livermore National Laboratory under Contract DE-AC52-07NA27344.
The assigned release number is LLNL-JRNL-843618.

\end{acknowledgments}

\section*{Data Availability}

\textcolor{black}{The data that support the findings of this study are available from the corresponding author upon request.}

\newpage

\bibliography{library}

\end{document}